\documentclass[
reprint,
 superscriptaddress,
 amsmath,amssymb,
prb,
floatfix,
]{revtex4-2}
\usepackage{graphicx}
\usepackage{dcolumn}
\usepackage{comment}
\usepackage{bm}
\usepackage{makecell}
\usepackage{tikz}
\usepackage{color}
\usepackage{braket}
\usepackage{array}
\usepackage{bbold}
\usepackage{cleveref}
\usepackage{tabularx}
\usepackage{booktabs}
\begin{document}

\title{Frequency-dependent electron-phonon coupling and vibrational responses in tight-binding and continuous Dirac models with nuclear velocity correction}

\author{Paolo Fachin}
\affiliation{Dipartimento di Fisica, Università di Roma La Sapienza, Roma, Italy}%
\author{Francesco Macheda}
\affiliation{Dipartimento di Fisica, Università di Roma La Sapienza, Roma, Italy}
\affiliation{Dipartimento di Scienze e Metodi dell’Ingegneria, Università di Modena e Reggio Emilia, Reggio Emilia, Italy}
\author{Paolo Barone}%
\affiliation{CNR-SPIN, Area della Ricerca di Tor Vergata, Via del Fosso del Cavaliere 100,
I-00133 Rome, Italy}%
\affiliation{Dipartimento di Fisica, Università di Roma La Sapienza, Roma, Italy}%
\author{Francesco Mauri}
\affiliation{Dipartimento di Fisica, Università di Roma La Sapienza, Roma, Italy}%

\newcommand{\citeSupp}[0]{Note1}

\begin{abstract}
The nuclear motion induces in the electronic atomic orbitals a nuclear-velocity-dependent phase (also known as electron-translation factor), which modifies the effective Hamiltonians constructed from localised atomic orbitals. In this work, using an Ehrenfest Lagrangian approach for the localised atomic orbitals (LCAO) and tight-binding methods, we determine, at any order in the nuclear velocity, the equations of motion and the vibrational responses within a linear response formalism, focusing on the tight-binding assessment of the Born effective charges and the force-constant matrix. The appearance of nuclear-velocity-dependent Peierls-like phases in the non-local part of the interactions restores the all-electron sum rules for frequency-dependent vibrational responses. In tight-binding models these corrections crucially modify the vibrational response from a qualitative point of view, also yielding contributions required to capture phenomena such as vibrational circular dichroism. We test these corrections in the tight-binding model for metallic gapped graphene - finding excellent agreement with \textit{ab initio} calculations - and for the topological time-reversal symmetry breaking Haldane model.   
\end{abstract}

\maketitle

\section*{Introduction}
Effective Hamiltonians, such as those constructed using a set of localised atomic orbitals as well as continuous low-energy models, are broadly used in condensed matter physics \cite{Grosso2013, Vanderbilt2018}. Tight-binding methods provide accurate descriptions of electronic properties of molecules and crystals through a reduced set of parameters, offering insights into physical properties with analytical results or at a significantly reduced computational cost \cite{harrison1989electronic}. Besides the description of the electronic band-structures, these models have been also employed in the recent decades for the assessment of the nuclear dynamics and vibrational response of the material. The inclusion of electron-phonon coupling in tight-binding and low-energy models provides reliable predictions of vibrational properties both in the adiabatic regime \cite{PhysRevB.65.235412,PhysRevLett.93.185503,doi:10.1143/JPSJ.75.084713,PhysRevB.84.035433,PhysRevB.90.125414,Bistoni2019,Villani2024,PhysRevB.110.L201405,Fachin2025, y1dn-m6pc, gnpt-5ffq} - including the chiral phonon phenomena \cite{bfll-sdrb, doi:10.7566/JPSJ.94.053601, PhysRevB.105.064303,PhysRevB.111.134414,doi:10.1021/acs.jpclett.6c00793,sato2025orbitalaccumulationinducedchiral,doi:10.7566/JPSJ.95.063705, PhysRevLett.134.206701} - and in non-adiabatic conditions \cite{PhysRevB.86.115439, Bistoni2019, Fachin2025, gnpt-5ffq, sellati2026lightinducedfaradayeffectdynamical}. In addition, this approach also enables the inclusion of many-body effects in the phonon spectra of graphene \cite{PhysRevB.111.075118}. The price to pay for such speed-ups is the emergence of non-local inter-atomic hoppings, not commuting with the position operator. Within the Born-Oppenheimer approximation, the effective electronic Hamiltonians are derived for a fixed nuclear configuration. Usually, even when nuclei are treated dynamically, the effective Hamiltonians are obtained by rigidly shifting the electronic orbital to the time-dependent nuclear positions. Nonetheless, this approach introduces discrepancies between the non-adiabatic frequency-dependent vibrational responses of the effective models and the all-electron Hamiltonian. This difference manifests in the frequency-dependent sum rules relating the Born effective charges and the force constant matrix to the frequency-dependent electromagnetic susceptibilities, as recently pointed out in the pseudopotential framework by Refs. \cite{PhysRevLett.128.095901, Marchese2024}.  These discrepancies are much larger in tight-binding methods where the sum rules are always vanishing \cite{Bistoni2019, Fachin2025}.  These issues arise from neglecting the effect of nuclear motion on the atomic-like orbitals, causing the appearance of nuclear-velocity dependent phases (also known as "electron-translation factors" in the literature) in the non-local part of the potentials \cite{PhysRevLett.136.196401, z497-65ks}. In tight-binding models the scale of the non-locality is the internuclear distance, much larger than in the non-local part of pseudopotential Hamiltonians, where it is limited to the atomic core region. Their use enables the recovery of the all-electron non-adiabatic vibrational response and dynamics in the pseudopotential framework \cite{PhysRevLett.136.196401, PAW_VI_2026, z497-65ks}. 

Moreover, the nuclear velocity-dependent phases make the Born-Oppenheimer ground state wave function complex valued, enabling non-zero electronic currents and magnetic dipole moments. Velocity-including atomic orbitals have been broadly used in the study of atomic collisions
\cite{Bates1958_electroncapture,LFErrea_1994, PhysRevA.50.418, PhysRevA.35.70,PhysRevA.85.012702,RJAllan_1985,PhysRevA.18.117, z497-65ks, PhysRevA.82.060701, PhysRevA.23.2301} and to assess the vibrational circular dichroism (VCD) of molecules and solids \cite{10.1063/1.462668,10.1063/1.445588,10.1063/1.4928578,2023_Ditler_Mattiat_Luber, doi:10.1021/ct400700c, doi:10.1021/acs.jctc.2c00006, doi:10.1021/acs.jpca.5c01344}. In addition, velocity-including atomic orbitals ensure Galilean invariance in the non-adiabatic couplings governing the time-dependent non-adiabatic dynamics. In this context, they have been used, to the linear order in the nuclear velocity, in time-dependent Hartree-Fock \cite{doi:10.1021/jp9906839, PhysRevA.62.022703, REYES2002441, doi:10.1021/jp505767b}, for localised atomic orbitals  \cite{10.1063/1.3665031, doi:10.1021/jz3006173, 10.1063/5.0160965}, in TDDFT linear response theory in Refs. \cite{10.1063/1.4906941,doi:10.1021/acs.jctc.5c01960}, as well as, to all orders in the nuclear velocity, in the pseudopotential PAW framework \cite{PAW_VI_2026}. Nevertheless, their use in the context of non-adiabatic vibrational properties remains sparse. For instance, velocity-including atomic orbitals have not been used yet for the assessment of the frequency-dependent electron-phonon coupling in tight-binding and low-energy continuous models. In these models, the frequency-dependent vibrational responses differ, not only quantitatively, but also qualitatively from the all-electron ones, requiring the development of a theory that overcomes this qualitative failure of the currently used tight-binding models. 

In this work, we incorporate the effect of the nuclear velocity on electronic atomic orbitals in localised atomic orbitals (LCAO), tight-binding and low-energy continuous Hamiltonians to derive the frequency-dependent vibrational responses satisfying the all-electron sum rules. Our derivations follow the same approach of Ref. \cite{PAW_VI_2026},  focused on the effect of nuclear velocity dependent phases in the first-principles non-adiabatic dynamics. We employ a semiclassical Ehrenfest Lagrangian approach, where nuclei are classical and electrons quantum mechanical particles \cite{Todorov_2001}.  Within this framework, we obtain the dynamics for both electrons and nuclei and, by using a linear response approach, we calculate the frequency-dependent vibrational responses, enabling the recovery of all-electron properties. From a physical viewpoint, we include the electronic inertia following a time-dependent nuclear displacement, quantified by the sum rules on vibrational responses, relating them to the frequency-dependent electronic susceptibilities \cite{PhysRevLett.128.095901, Marchese2024,PhysRevLett.129.185902, PhysRevB.107.094308, PhysRevB.110.094306, PhysRevB.111.094307}, that was completely lacking in tight-binding models constructed neglecting the nuclear velocity effects \cite{Bistoni2019, Fachin2025}.  Therefore, our work provides not only quantitative corrections to the frequency-dependent vibrational responses, but fundamental qualitative changes, accounting for features that were absent in standard tight-binding approximations for lattice dynamics.

The paper is divided into three main parts: a general theoretical derivation of the effective Hamiltonians in the presence of nuclear motion in Section \ref{sec:effective_Hamiltonians} that follows the same approach used for the pseudopotential framework in Ref. \cite{PAW_VI_2026}; a general description of the frequency-dependent vibrational responses for tight-binding methods in Section \ref{sec:vibrationa_responses}; the application to gapped graphene and to the Haldane model in Sec. \ref{sec:appl}. In detail, the first section is organized as follows: we review the velocity-including atomic orbitals for an isolated nucleus in motion in Sec. \ref{sec:I}; the adiabatic LCAO and tight-binding methods in Sec. \ref{sec:adiabatic LCAO}.  In Sec. \ref{sec:II} we derive the semiclassical Lagrangian for the LCAO method, obtaining the effective Hamiltonian and the equations of motions for the nuclei and the electrons. In Sec. \ref{sec:TB} we adopt the orthogonal tight-binding approximation on the LCAO Lagrangian, determining explicitly the tight-binding Hamiltonian and the linear response to the nuclear displacement. The tight-binding results are used to determine the effect of the nuclear velocity dependent phases in the low-energy Dirac models in Section \ref{sec:kp_models}. In these models, the correction has a simple expression, enabling a clear physical interpretation in terms of the band velocity.  In the second part, we derive, for these models, in Section \ref{sec:vibrationa_responses}, the explicit expressions for the vibrational responses, focusing on the Born effective charges and the force constant matrix. In the final part, in Sec. \ref{sec:appl} we study the corrections to the Born effective charges and to the force constant matrix in metallic gapped graphene and in the topologically non-trivial time-reversal symmetry breaking Haldane model when the vibrational excitations resonate with the electronic interband transitions.
Our conclusions are drawn in Sec. \ref{sec:concl} as well as summarised in Tables \ref{tab:summary_eph} and \ref{tab:Dirac_Hamiltonian_electron-phonon}, where we report the effective Hamiltonians accounting for the nuclear velocities effects as well as the explicit tight-binding and low-energy frequency-dependent nuclear displacement derivatives. 

\section{Effective Hamiltonians with nuclear velocity effects}\label{sec:effective_Hamiltonians}
\subsection{Velocity-including atomic orbitals}
\label{sec:I}
In this Section, we define a localised atomic orbital basis for the case of moving nuclei, summarising the detailed derivation given in Ref. \cite{PAW_VI_2026}.
 
Consider an isolated atom $s$ with the nucleus at rest, which, without loss of generality, is located at the origin. The electrons are described by a single particle all-electron Hamiltonian 
\begin{equation}
\hat{H}_{\mathbf{0}}^{\mathrm{AE}}(\hat{\mathbf{r}},\hat{\mathbf{p}})=\frac{\hat{\mathbf{p}}^2}{2m}+V_s(\hat{\mathbf{r}}).
\label{eq:all-electron_hamiltonian_static}
\end{equation}
where the hat $\hat{}$ denotes operators, $V_s(\hat{\mathbf{r}})$ is the effective potential acting on the electron, the subscript $s$ indicates the dependence of the potential on the atomic properties of $s$. The Hamiltonian is diagonalised by the atomic orbitals $\ket{\phi^{\rm \mathbf{0}}_{si}}$ with energy $E_{si}$, where $i$ indicates the electronic quantum number and the superscript $\mathbf{0}$ indicates that the orbitals $\ket{\phi^{\rm \mathbf{0}}_{si}}$ are solutions for the atom at rest at the origin,
\begin{equation}
    \hat{H}_{\mathbf{0}}^{\mathrm{AE}}(\hat{\mathbf{r}},\hat{\mathbf{p}})\ket{\phi^{\rm \mathbf{0}}_{si}}=E_{si}\ket{\phi^{\rm \mathbf{0}}_{si}}.
\end{equation}
Usually, localised atomic orbital bases are constructed using $\ket{\phi^{\rm \mathbf{0}}_{si}}$ orbitals, translated to the positions of atoms of the system
\begin{equation}
    \ket{\phi^{\mathbf{R}_{s}}_{si}}=\hat{T}_{\mathbf{R}_{s}}\ket{\phi^{\mathbf{0}}_{si}},
\end{equation}
where the translation operator  $\hat{T}_{\mathbf{R}_{s}}$ acts on the position eigenstates as $\hat{T}_{\mathbf{R}_s}\ket{\mathbf{r}}=\ket{\mathbf{r}+\mathbf{R}_s}$. These orbitals form the basis set $\{\ket{\phi^{\mathbf{R}_{s}}_{si}}\}_{si}$.  As discussed in detail in Ref. \cite{PAW_VI_2026}, the electronic atomic orbital basis that accounts for the nuclear motion is defined as
\begin{align}
&\ket{\phi^{\dot{\mathbf{R}}_s,\mathbf{R}_{s}}_{si}}=e^{i\alpha_{s}(\hat{\mathbf{r}})}\ket{\phi^{\mathbf{R}_{s}(t)}_{si}},\label{eq:vi_atomic_orbital}\\
&\alpha_{s}(\hat{\mathbf{r}})=\frac{m}{\hbar}\dot{\mathbf{R}}_s(t)\cdot(\hat{\mathbf{r}}-\mathbf{R}_s(t)). \nonumber
\end{align} 
In the superscript $\dot{\mathbf{R}}_s,\mathbf{R}_{s}$ we omit the temporal dependence for brevity since it is clear from the context. The velocity-including basis set $\{{\ket{\phi^{\dot{\mathbf{R}}_s,\mathbf{R}_{s}}_{si}}}\}_{si}$ depends on time instantaneously through the nuclear position $\mathbf{R}_s(t)$ and velocity $\dot{\mathbf{R}}_s(t)$. 
The temporal derivative of the states is 
\begin{equation}
    \begin{aligned}
&\frac{d\ket{\phi^{\dot{\mathbf{R}}_s,\mathbf{R}_{s}}_{si}}}{dt}=e^{i\alpha_s(\hat{\mathbf{r}})}\frac{i}{\hbar}\Bigg(m\ddot{\mathbf{R}}_s(t)\cdot(\hat{\mathbf{r}}-\mathbf{R}_s(t))\\&-\dot{\mathbf{R}}_s(t)\cdot \hat{\mathbf{p}}-m|\dot{\mathbf{R}}_s(t)|^2\Bigg) 
\ket{\phi^{\mathbf{R}_{s}(t)}_{si}},\label{eq:derivative_vi_atomic_orbital}\\
\end{aligned} 
\end{equation}
and by using the commutation relation of the position and momentum operators, 
\begin{equation}
    \begin{aligned}
\frac{d\ket{\phi^{\dot{\mathbf{R}}_s,\mathbf{R}_{s}}_{si}}}{dt}=&\frac{i}{\hbar}\Bigg(m\ddot{\mathbf{R}}_s(t)\cdot(\hat{\mathbf{r}}-\mathbf{R}_s(t))\\&-\dot{\mathbf{R}}_s(t)\cdot \hat{\mathbf{p}}\Bigg) 
\ket{\phi^{\dot{\mathbf{R}}_s,\mathbf{R}_{s}}_{si}}.\label{eq:derivative_vi_atomic_orbital_v2}
\end{aligned} 
\end{equation}
As discussed in Ref. \cite{PAW_VI_2026}, the orbitals in Eq. \eqref{eq:vi_atomic_orbital} are a solution of the Schrödinger equation with the potential centred on the moving nucleus if the nuclear acceleration term, causing a Stark-like effect, is neglected.  Therefore, in the construction of the basis, we are excluding the mixing, due to the Stark effect, of the occupied states with orbitals that are not included in the basis set. These contributions are small because of the large energy difference between the occupied states and those excluded from the basis set, becoming increasingly smaller as the basis set is enlarged. 

The notation used to indicate the atomic orbitals is summarised in Table \ref{tab:atomic_orbitals_legend}.

\begin{table}[h!]
    \centering
    \renewcommand{\arraystretch}{2} 
    \begin{tabular}{l|l}
    &  Atomic orbital for level $i$ of atom $s$\\
    \hline
    $\ket{\phi^{\mathbf{0}}_{si}}$   
        & for the atom centred at the origin \\
    \hline
    $\ket{\phi^{\mathbf{R}_{s}}_{si}}$   
        & translated to the fixed atomic equilibrium position \\
    \hline
    $\ket{\phi^{\mathbf{R}_{s}(t)}_{si}}$   
        & translated to the time-dependent atomic position \\
    \hline
    $\ket{\phi^{\dot{\mathbf{R}}_s,\mathbf{R}_{s}}_{si}}$   
        & $\ket{\phi^{\mathbf{R}_{s}(t)}_{si}}$ times nuclear velocity-dependent phase \\
    \end{tabular}
    \caption{Summary of the notation used for the different kinds of atomic orbitals throughout the paper.}
    \label{tab:atomic_orbitals_legend}
\end{table}

\subsection{Adiabatic LCAO and tight-binding}\label{sec:adiabatic LCAO}
\subsubsection{LCAO}
In the adiabatic LCAO method describing a molecule or a crystal, the $I$-th eigenstate single-particle all-electron wavefunction $\ket{\psi_I}$ is expanded on a set of localised atomic orbitals rigidly centred in the static nuclear positions, $\displaystyle \mathbf{R}_b$, $\displaystyle \{\ket{\phi^{\mathbf{R}_b}_{bi}}\}$ as $ \displaystyle \ket{\psi_I}=\sum_{b,i} c^I_{bi}\ket{\phi^{\mathbf{R}_b}_{bi}}$, where $c^I_{bi}$ are the expansion coefficients of the $I$-th state on the $i$-th atomic orbital of the $b$ nucleus \cite{Grosso2013}. The all-electron wavefunctions are orthonormal $\braket{\psi_J|\psi_I}=\delta_{JI}$. By defining the overlap matrix
\begin{align}
\mathcal{S}^{\rm R}_{bj,b'i}=\braket{\phi^{\mathbf{R}_b}_{bj}|\phi^{\mathbf{R}_{b'}}_{b'i}}
\label{eq:S_lcao_static}
\end{align}
and the LCAO Hamiltonian
\begin{equation}
\begin{aligned}
&\mathcal{H}^{\rm LCAO}_{bj,b'i}=\braket{\phi^{\mathbf{R}_b}_{bj}|H^{\mathrm{AE}}|\phi^{\mathbf{R}_{b'}}_{b'i}},
\end{aligned}
\label{eq:LCAO_Hamiltonian_static}
\end{equation}
the energies and the eigenvectors are obtained as
\begin{equation}
    \begin{aligned}
\sum_{b',i} \mathcal{H}^{\mathrm{LCAO}}_{bj,b'i} c^I_{b'i}=E_I\sum_{b',i} \mathcal{S}^{\rm R}_{bj,b'i} c^I_{b'i}.
\label{eq:SCHR_LCAO_static}
\end{aligned}
\end{equation}
In the equations of motion for the nuclei, we distinguish a Hellmann-Feynman $\mathbf{F}_s^{\mathrm{HF}}$ and Pulay $\mathbf{F}_s^{\mathrm{P}}$ forces
\begin{equation}
    M_s \ddot{\mathbf{R}}_s(t)=\mathbf{F}_s^{\mathrm{HF}}+\mathbf{F}_s^{\mathrm{P}},
    \label{eq:Ehrenfest_ions_2}
\end{equation}
where, in detail, 
\begin{equation}
    \begin{aligned}
        \mathbf{F}_s^{\mathrm{HF}}=-\sum_{I=1}^{N_{\rm el}}\sum_{bj,b'i}\left(c^{I}_{bj}\right)^*\frac{\partial \mathcal{H}^{\mathrm{LCAO}}_{bj,b'i} }{\partial \mathbf{R}_s}c^I_{b'i},
    \end{aligned}
    \label{eq:HF_force}
\end{equation}
\begin{equation}
    \begin{aligned}
   &\mathbf{F}_s^{\mathrm{P}}= \sum_{I=1}^{N_{\rm el}}\sum_{bj,b'i}E_I\left(c^{I}_{bj}\right)^*\frac{\partial \mathcal{S}^{\rm R}_{bj,b'i}}{\partial \mathbf{R}_s}c^I_{b'i}.
    \end{aligned}
    \label{eq:pulay_forces_adiabatic_LCAO}
\end{equation}
\subsubsection{Orthogonal tight-binding}
Assuming that the orbitals in the basis are orthogonal  $\mathcal{S}^{\rm R}_{bj,b'i}=\braket{\phi^{\mathbf{R}_{b}}_{bj}|\phi^{\mathbf{R}_{b'}}_{b'i}}=\delta_{bb'}\delta_{ij}$ and defining the on-site energy and the hopping, respectively, as
\begin{equation}
\Delta_{bi}=\braket{\phi^{\mathbf{R}_b(t)}_{bi}|H^{\mathrm{AE}}|\phi^{\mathbf{R}_b(t)}_{bi}}
\end{equation}
and
\begin{equation}
t^0_{bj,b'i}=\bra{\phi^{\mathbf{R}_b}_{bj}}\frac{\hat{\mathbf{p}}^2}{2m}+V(\hat{\mathbf{r}})\ket{\phi^{\mathbf{R}_{b'}}_{b'i}},\quad b\neq b',
\label{eq:static_hopping}
\end{equation}
we obtain the tight-binding Hamiltonian
\begin{equation}
  \mathcal{H}_{bj,b'i}^{\rm TB}=t^0_{bj,b'i}(1-\delta_{bb'})+\delta_{bb'}\delta_{ij}\Delta_{bi}.
\label{eq:TB_hamiltonian_adiabatic}
\end{equation}
From Eq. \eqref{eq:SCHR_LCAO_static}, it follows that the diagonalisation of the tight-binding Hamiltonian matrix yields the energies and the related eigenvectors. The nuclear dynamics is governed by the Hellmann-Feynman contribution with the tight-binding Hamiltonian. 
\subsection{Lagrangian approach for the velocity-including LCAO method}
\label{sec:II}
In this Section, we derive the effective Hamiltonian and the non-adiabatic equations of motion with the Lagrangian approach, following the same procedure of Ref. \cite{PAW_VI_2026}.
Consider a system of classical nuclei, located at the positions $\{\mathbf{R}_s(t)\}$, and quantum electrons.  The all-electron single-particle mean-field Hamiltonian $\hat{H}$ presents an effective self-consistent local potential $V(\hat{\mathbf{r}})$, obtained with a density functional theory (DFT) local or semi-local approximation for the exchange-correlation functional. Therefore, the single-particle Hamiltonian of the electron interacting with many nuclei is
\begin{equation}
\hat{H}^{\mathrm{AE}}(\hat{\mathbf{r}},\hat{\mathbf{p}};\{\mathbf{R}_s(t)\})=\frac{\hat{\mathbf{p}}^2}{2m}+V(\hat{\mathbf{r}}).
\label{eq:HAEsingle}
\end{equation}
At zero temperature, the system of electrons and nuclei can be described through the real-valued Ehrenfest Lagrangian \cite{Todorov_2001}, depending on the independent variables $\mathbf{q}=\left(\{\mathbf{R}_s(t)\}_s, \{\ket{\psi_I(t)}\}_I, \{\bra{\psi_I(t)}\}_I\right)$,
\begin{equation}
\begin{aligned}
    & \mathcal{L}^{\rm AE}(\mathbf{q})=\sum_s\frac{M_s|\dot{\mathbf{R}}_s(t)|^2}{2}-\sum_{I=1}^{N_{\rm el}}\Bigg(\braket{\psi_I(t)|\hat{H}^{\mathrm{AE}}|\psi_I(t)}\\&-\frac{i\hbar}{2}\left(\bra{\psi_I(t)} \frac{d\ket{\psi_I(t)}}{dt}- \frac{d\bra{\psi_I(t)}}{dt}\ket{\psi_I(t)}\right)\Bigg)
\end{aligned}
    \label{eq:lagrangian_general}
\end{equation} 
where $\ket{\psi_I(t)}$ are the solutions of Eq. \eqref{eq:HAEsingle}, and $N_{\rm el}$ is the number of electrons. 
We do not treat the finite temperature case directly in the Lagrangian, but we will recover it via the standard expressions for the linear response in the following Sections.

The all-electron wavefunction is expanded in terms of the nuclear-velocity including atomic orbitals given in Eq. \eqref{eq:vi_atomic_orbital} as $\ket{\psi_I(t)}=\sum_{bi} c^I_{bi}(t) \ket{\phi^{\dot{\mathbf{R}}_b,\mathbf{R}_{b}}_{bi}}$. Defining the overlap matrix as
 \begin{align}
\mathcal{S}^{\rm R, \dot{R}}_{bj,b'i}=\braket{\phi^{\dot{\mathbf{R}}_b,\mathbf{R}_{b}}_{bj}|\phi^{\dot{\mathbf{R}}_{b'},\mathbf{R}_{b'}}_{b'i}},
\label{eq:S_lcao_R}
\end{align}
we obtain the LCAO Lagrangian depending on the nuclear positions $\{\mathbf{R}_s(t)\}$ and the electronic coefficients $\{c^{I}_{bi}(t)\},\{\left(c^{I}_{bi}(t)\right)^*\}$  as independent variables

    \begin{align}
&\mathcal{L}_{\mathrm{R,\dot{R}}}^{\mathrm{LCAO}}= \sum_s\frac{M_s|\dot{\mathbf{R}}_s(t)|^2}{2}
        -\sum_{I=1}^{N_{\rm el}}\sum_{b'j,bi}\Bigg[\left(c^{I}_{bj}(t)\right)^*\mathcal{H}^{\mathrm{R,\dot{R}}}_{bj,b'i}c^I_{b'i}(t)\nonumber\\&-\frac{i\hbar}{2} \Bigg(\left(c^{I}_{bj}(t)\right)^*\mathcal{S}^{\mathrm{R,\dot{R}}}_{bj,b'i}\dot{c}^I_{b'i}(t)-\dot{c}^{I,*}_{bj}(t)\mathcal{S}^{\mathrm{R,\dot{R}}}_{bj,b'i}c^I_{b'i}(t)\Bigg)\Bigg].
    \label{eq:LCAO_lagrangian}
\end{align}
where
\begin{align}
\mathcal{H}^{\mathrm{R,\dot{R}}}_{bj,b'i}=\mathcal{H}^{\mathrm{LCAO}}_{bj,b'i}
+\mathcal{M}_{bj,b'i}^{\rm R, \dot{R}}, \label{eq:LCAO_Hamiltonian_R}
\end{align}
\begin{align}
&\mathcal{H}^{\mathrm{LCAO}}_{bj,b'i}=\braket{\phi^{\dot{\mathbf{R}}_b,\mathbf{R}_{b}}_{bj}|H^{\mathrm{AE}}|\phi^{\dot{\mathbf{R}}_{b'},\mathbf{R}_{b'}}_{b'i}},\\
&\mathcal{M}_{bj,b'i}^{\rm R, \dot{R}}=\frac{i\hbar}{2} \Bigg(\frac{d\bra{\phi^{\dot{\mathbf{R}}_b,\mathbf{R}_{b}}_{bj}}}{dt}\ket{\phi^{\dot{\mathbf{R}}_{b'},\mathbf{R}_{b'}}_{b'i} }- \bra{\phi^{\dot{\mathbf{R}}_b,\mathbf{R}_{b}}_{bj}}\frac{d\ket{\phi^{\dot{\mathbf{R}}_{b'},\mathbf{R}_{b'}}_{b'i}}}{dt}\Bigg).
\end{align}

The LCAO Lagrangian is analogous to the PAW one given in Eq. (53) of Ref. \cite{PAW_VI_2026}, replacing the coefficients of the LCAO expansion with the pseudowavefunctions. The atomic orbital basis plays a similar role to the PAW transformation operator. Therefore, also for equations of motions this analogy holds, as detailed in the following. 
With some algebraic manipulations based on the derivative of the atomic orbitals,  shown in Appendix \ref{app:1}, we obtain
\begin{widetext}
\begin{align}
 &\mathcal{H}^{\mathrm{R,\dot{R}}}_{bj,b'i}=           \bra{\phi^{\mathbf{R}_{b}(t)}_{bj}}\Bigg(e^{i(\alpha_{b'}(\hat{\mathbf{r}})-\alpha_{b}(\hat{\mathbf{r}}))}\left(\frac{\hat{\mathbf{p}}^2}{4m}+\frac{V(\hat{\mathbf{r}})}{2}\right)
   +\left(\frac{\hat{\mathbf{p}}^2}{4m}+\frac{V(\hat{\mathbf{r}})}{2}\right)e^{i(\alpha_{b'}(\hat{\mathbf{r}})-\alpha_{b}(\hat{\mathbf{r}}))}\Bigg)\ket{\phi^{\mathbf{R}_{b'}(t)}_{b'i}}\nonumber\\
  &+ \frac{1}{2}\bra{\phi^{\mathbf{R}_{b}(t)}_{bj}}\Bigg(m \ddot{\mathbf{R}}_{b'}(t)\cdot\left(\hat{\mathbf{r}}-\mathbf{R}_{b'}(t)\right)+m\ddot{\mathbf{R}}_{b}(t)\cdot\left(\hat{\mathbf{r}}-\mathbf{R}_{b}(t)\right)
-\frac{m}{2}\Big(|\dot{\mathbf{R}}_{b'}(t)|^2+|\dot{\mathbf{R}}_{b}(t)|^2\Big)\Bigg)e^{i(\alpha_{b'}(\hat{\mathbf{r}})-\alpha_{b}(\hat{\mathbf{r}}))}\ket{\phi^{\mathbf{R}_{b'}(t)}_{b'i}}.
\label{eq:HLCAORDOT}
\end{align}

\end{widetext}
The equations of motion for the nuclear and electronic variables are determined by imposing the action $\mathcal{A}=1/T\int_0^T \mathcal{L}dt$ to be stationary for a variation of an independent variable of the Lagrangian, i.e.  $\frac{\delta \mathcal{A}}{\delta q_i}=0$.  The electronic dynamics is obtained using the standard Euler-Lagrange equations,
\begin{equation}
    \begin{aligned}
\sum_{b',i}i\hbar \mathcal{S}^{\mathrm{R,\dot{R}}}_{bj,b'i}\dot{c}^I_{b'i}(t)=\\\sum_{b',i}\Bigg(\mathcal{H}^{\mathrm{LCAO}}_{bj,b'i}-i\hbar\bra{\phi^{\dot{\mathbf{R}}_b,\mathbf{R}_{b}}_{bj}}\frac{d\ket{\phi_{b'i}^{\dot{\mathbf{R}}_{b'},\mathbf{R}_{b'}}}}{dt}\Bigg) c^I_{b'i}(t).
\label{eq:SCHR_LCAO_translating}
\end{aligned}
\end{equation}
The effective Hamiltonian governing the temporal evolution is not Hermitian, if the overlap matrix is time-dependent, still guaranteeing the norm-conservation.  If the overlap matrix is diagonal, the effective Hamiltonian governing the temporal evolution of the system coincides with the Hermitian $\mathcal{H}^{\mathrm{R,\dot{R}}}$. The same arguments hold also in the PAW case, where the electronic dynamics in Eq. (63) of Ref. \cite{PAW_VI_2026} has a clear correspondence with the proper replacement, discussed for the Lagrangian above.

Conversely, the nuclear-velocity dependence of the atomic orbitals causes a nuclear-acceleration dependence in the effective model's Lagrangian. Under the assumption that the highest order temporal derivative is the second order one, the Euler-Lagrange equations are generalised as \cite{woodard2015theoremostrogradsky, ostrogradsky1850memoires}
\begin{equation}
    \frac{\partial \mathcal{L}(\mathbf{q},\dot{\mathbf{q}},\ddot{\mathbf{q}})}{\partial q_i}-\frac{d}{dt}\frac{\partial \mathcal{L}(\mathbf{q},\dot{\mathbf{q}},\ddot{\mathbf{q}})}{\partial \dot{q}_i}+\frac{d^2}{dt^2}\frac{\partial \mathcal{L}(\mathbf{q},\dot{\mathbf{q}},\ddot{\mathbf{q}})}{\partial \ddot{q}_i}=0.
    \label{eq:Euler-Lagrange_general}
\end{equation}
Specifically, the equations for the nuclear system are obtained by setting $\mathbf{q}_i \to \mathbf{R}_s$. The linear dependence of the Lagrangian on the acceleration forbids the presence of nuclear position time-derivatives of order higher than the second in the equations of motion, avoiding the issues of Ostrogradsky’s instabilities \cite{ostrogradsky1850memoires, woodard2015theoremostrogradsky, PhysRevD.91.085009}. We distinguish different contributions to the forces governing the nuclear dynamics
\begin{equation}
    M_s \ddot{\mathbf{R}}_s(t)=\mathbf{F}_s^{\mathrm{HF-R}}+\mathbf{F}_s^{\mathrm{HF-\dot{R}}}+\mathbf{F}_s^{\mathrm{HF-\ddot{R}}}+\mathbf{F}_s^{\mathrm{P}}+\mathbf{F}_s^{\mathrm{P-\dot{R}}},
    \label{eq:Ehrenfest_ions}
\end{equation}
where $\mathbf{F}_s^{\mathrm{HF-R}},\mathbf{F}_s^{\mathrm{HF-\dot{R}}}\text{ and }\mathbf{F}_s^{\mathrm{HF-\ddot{R}}}$ are the Hellmann-Feynman-like contributions to the force originating from nuclear position, velocity and acceleration derivatives of the Lagrangian; $\mathbf{F}_s^{\mathrm{P}}$ and $\mathbf{F}_s^{\mathrm{P-\dot{R}}}$ are the Pulay forces \cite{Pulay_1969} originating from the non-orthogonality of the basis and its generalisation to the nuclear velocity derivatives. In detail, their expressions are 
\begin{equation}
    \begin{aligned}
        \mathbf{F}_s^{\mathrm{HF-R}}=-\sum_{I=1}^{N_{\rm el}}\sum_{bj,b'i}\left(c^{I}_{bj}(t)\right)^*\frac{\partial \mathcal{H}^{\mathrm{R,\dot{R}}}_{bj,b'i} }{\partial \mathbf{R}_s}c^I_{b'i}(t),
    \end{aligned}
    \label{eq:HF_force}
\end{equation}

    \begin{align}
        \mathbf{F}_s^{\mathrm{HF-\dot{R}}}=\frac{d}{dt}\sum_{I=1}^{N_{\rm el}}\sum_{bj,b'i}\left(c^{I}_{bj}(t)\right)^*\frac{\partial \mathcal{H}^{\mathrm{R,\dot{R}}}_{bj,b'i} }{\partial \dot{\mathbf{R}}_s}c^I_{b'i}(t),  \label{eq:HF_force_dotR}
    \end{align}
  
    \begin{align}
        \mathbf{F}_s^{\mathrm{HF-\ddot{R}}}=-\frac{d^2}{dt^2}\sum_{I=1}^{N_{\rm el}}\sum_{bj,b'i}\left(c^{I}_{bj}(t)\right)^*\frac{\partial \mathcal{H}^{\mathrm{R,\dot{R}}}_{bj,b'i}}{\partial \ddot{\mathbf{R}}_s}c^I_{b'i}(t).    \label{eq:HF_force_ddotR}
    \end{align}

    \begin{align}
   &\mathbf{F}_s^{\mathrm{P}}= \frac{i\hbar}{2}\sum_{I=1}^{N_{\rm el}}\sum_{bj,b'i}\Bigg(\left(c^{I}_{bj}(t)\right)^*\frac{\partial \mathcal{S}^{\mathrm{R,\dot{R}}}_{bj,b'i}}{\partial \mathbf{R}_s}\dot{c}^I_{b'i}(t)\nonumber\\
   &\qquad\qquad\quad-\left(\dot{c}^{I}_{bj}(t)\right)^*\frac{\partial \mathcal{S}^{\mathrm{R,\dot{R}}}_{bj,b'i}}{\partial \mathbf{R}_s}c^I_{b'i}(t)\Bigg),  \label{eq:pulay_forces}
    \end{align}

    \begin{align}
    &\mathbf{F}_s^{\mathrm{P-\dot{R}}}= -\frac{i\hbar}{2}\sum_{I=1}^{N_{\rm el}}\sum_{bj,b'i}\frac{d}{dt}\Bigg( \left(c^{I}_{bj}(t)\right)^*\frac{\partial \mathcal{S}^{\mathrm{R,\dot{R}}}_{bj,b'i}}{\partial \dot{\mathbf{R}}_s}\dot{c}^I_{b'i}(t)\nonumber\\&
    \qquad \qquad -\left(\dot{c}^{I}_{bj}(t)\right)^*\frac{\partial \mathcal{S}^{\mathrm{R,\dot{R}}}_{bj,b'i}}{\partial \dot{\mathbf{R}}_s}c^I_{b'i}(t)
    \Bigg).\label{eq:pulay_forces_dotR}
    \end{align}
The above equations are analogous to the forces governing the nuclear dynamics in the PAW case, Eqs.  (66),(67),(70) and (71) of Ref. \cite{PAW_VI_2026}, upon replacement of PAW pseudo-wavefunction with LCAO coefficients. The same holds also for the conserved energy that is derived below (see Eq. (73) of  Ref. \cite{PAW_VI_2026} for a comparison). Using the Ostrogradsky formulation, as appropriate due to the (linear) dependence of the Lagrangian on the nuclear acceleration, the conserved energy can be expressed as
 \begin{widetext}
\begin{equation}
    \begin{aligned}
E_{\mathrm{R,\dot{R}}}=&\sum_s\frac{M_s|\dot{\mathbf{R}}_s(t)|^2}{2}+\sum_{I=1}^{N_{\rm el}}\sum_{bj,b'i}\Bigg\{\left(c^{I}_{bj}(t)\right)^*\mathcal{H}^{\mathrm{R,\dot{R}}}_{bj,b'i}c^I_{b'i}(t)-\sum_s\Bigg[\left(c^{I}_{bj}(t)\right)^*\left(\frac{\partial \mathcal{H}^{\mathrm{R,\dot{R}}}_{bj,b'i}}{\partial \dot{\mathbf{R}}_s } \cdot \dot{\mathbf{R}}_s+\frac{\partial \mathcal{H}^{\mathrm{R,\dot{R}}}_{bj,b'i}}{\partial \ddot{\mathbf{R}}_s }\cdot \ddot{\mathbf{R}}_s (t)\right)c^I_{b'i}(t)\\
     &-\frac{d}{dt}\left(\left(c^{I}_{bj}(t)\right)^*\frac{\partial \mathcal{H}^{\mathrm{R,\dot{R}}}_{bj,b'i}}{\partial \ddot{\mathbf{R}}_s }c^I_{b'i}(t)\right)\cdot \dot{\mathbf{R}}_s(t) +\frac{i\hbar}{2}\left(\dot{c}^{I,*}_{bj}(t)\frac{\partial \mathcal{S}^{\mathrm{R,\dot{R}}}_{bj,b'i}}{\partial \dot{\mathbf{R}}_s}c^I_{b'i}(t)-\left(c^{I}_{bj}(t)\right)^*\frac{\partial \mathcal{S}^{\mathrm{R,\dot{R}}}_{bj,b'i}}{\partial \dot{\mathbf{R}}_s}\dot{c}^I_{b'i}(t)\right)\cdot \dot{\mathbf{R}}_s(t)\Bigg]\Bigg\}.
     \label{eq:conserved_LCAO-DOTR}
 \end{aligned}
\end{equation}     
\end{widetext}
As for the equations of motion, the linearity of the Lagrangian in the nuclear acceleration forbids the presence of derivatives of the position of order higher than the second in the conserved energy.

\subsection{Velocity-including tight-binding}
\label{sec:TB}
In this Section, we evaluate the LCAO Lagrangian of Eq. \eqref{eq:LCAO_lagrangian} for velocity-including atomic orbitals within the tight-binding approximation. We assume an orthonormal basis, implying a diagonal time-independent overlap matrix $\mathcal{S}^{\mathrm{R,\dot{R}}}_{bj,b'i}=\braket{\phi^{\dot{\mathbf{R}}_{b},\mathbf{R}_{b}}_{bj}|\phi^{\dot{\mathbf{R}}_{b'},\mathbf{R}_{b'}}_{b'i}}=\delta_{bb'}\delta_{ij}$. Our aim is to obtain an expression for the tight-binding hopping and the on-site energy in the basis of the velocity-including atomic orbitals in relation to the hoppings and on-site energies in the basis of the orbitals for fixed nuclei. To evaluate the localised atomic-orbital Hamiltonian within the tight-binding approximation, we first consider the on-site energy contribution, where both orbitals are on the same site, and then the hopping terms connecting orbitals on different sites.  

\subsubsection{On-site energy}
If the two atomic orbitals belong to the same site, the phases cancel and the Hamiltonian of Eq. \eqref{eq:HLCAORDOT} reduces to
\begin{equation}
    \begin{aligned}
   \mathcal{H}^{\mathrm{R,\dot{R}}}_{bj,bi}=&\bra{\phi^{\mathbf{R}_{b}(t)}_{bj}}\frac{\hat{\mathbf{p}}^2}{2m}+V(\hat{\mathbf{r}})-\frac{m|\dot{\mathbf{R}}_{b}(t)|^2}{2}\\&+m\ddot{\mathbf{R}}_{b}(t)\cdot (\hat{\mathbf{r}}-\mathbf{R}_{b}(t))\ket{\phi^{\mathbf{R}_b(t)}_{bi}}.
    \end{aligned}
\end{equation}
The first two terms give the same on-site energy as for the atomic orbitals at fixed nuclei, $\Delta_{bi}=\braket{\phi^{\mathbf{R}_b(t)}_{bi}|H^{\mathrm{AE}}|\phi^{\mathbf{R}_b(t)}_{bi}}$. Assuming that the position operator is diagonal on the basis of the localised orbitals 
\begin{equation}
\hat{\mathbf{r}}\ket{\phi^{\mathbf{R}_b(t)}_{bi}}=\mathbf{R}_{b}(t)\ket{\phi^{\mathbf{R}_b(t)}_{bi}},
    \label{eq:diagonal_pos_operator}
\end{equation}
the nuclear acceleration term vanishes, yielding the following tight-binding Hamiltonian
\begin{equation}
  \mathcal{H}^{\mathrm{TB}}_{bj,bi}=\left(\Delta_{bi}-\frac{m|\dot{\mathbf{R}}_{b}(t)|^2}{2}\right)\delta_{ij}.
  \label{eq:TB_onsite_VI}
\end{equation}
The additional term, appearing for the moving nucleus, corresponds to the kinetic energy of the electron in the frame co-moving with the nucleus. The negative sign correspond to a positive contribution to the kinetic energy of the atom in the Lagrangian.

\subsubsection{Hopping} 
Hopping terms connect orbitals located on different sites. For atomic orbitals at fixed nuclei, the hopping is defined in Eq. \eqref{eq:static_hopping}, whereas, in the presence of nuclear motion, we define the hopping $t_{bj,b'i}$, depending on the positions of the moving nuclei, as
    \begin{align}
    &t_{bj,b'i}=\bra{\phi^{\mathbf{R}_{b}(t)}_{bj}} \frac{\hat{\mathbf{p}}^2}{2m}+V(\hat{\mathbf{r}}) \ket{\phi^{\mathbf{R}_{b'}(t)}_{b'i}},\quad b\neq b'. \label{eq:thopp}
\end{align}
For moving nuclei, the atom-centred contributions to the potential $V(\hat{\mathbf{r}}) $ originates from the implicit dependence of $V(\hat{\mathbf{r}})$ upon $\mathbf{R}_{s}(t)$.  When the nuclear positions are at the equilibrium configuration, $t_{bj,b'i}$ coincides with $t^0_{bj,b'i}$.  Under the assumption of an orthogonal basis set and of a diagonal position operator according to Eq. \eqref{eq:diagonal_pos_operator}, the Hamiltonian of Eq. \eqref{eq:HLCAORDOT} reduces to
\begin{align}   &\mathcal{H}^{\mathrm{R,\dot{R}}}_{bj,b'i}=\bra{\phi^{\mathbf{R}_{b}(t)}_{bj}}\Bigg(e^{i(\alpha_{b'}(\hat{\mathbf{r}})-\alpha_{b}(\hat{\mathbf{r}}))}\left(\frac{\hat{\mathbf{p}}^2}{4m}+\frac{V(\hat{\mathbf{r}})}{2}\right)\nonumber\\&
   +\left(\frac{\hat{\mathbf{p}}^2}{4m}+\frac{V(\hat{\mathbf{r}})}{2}\right)e^{i(\alpha_{b'}(\hat{\mathbf{r}})-\alpha_{b}(\hat{\mathbf{r}}))}\Bigg)\ket{\phi^{\mathbf{R}_{b'}(t)}_{b'i}}.
   \label{eq:hopping_tight_binding_general_standard}
    \end{align}
By observing that
\begin{align}
    &e^{i\alpha_{b'}(\hat{\mathbf{r}})}\ket{\phi^{\mathbf{R}_{b}(t)}_{bj}}=e^{\frac{i}{\hbar}m\dot{\mathbf{R}}_{b'}(t) \cdot (\mathbf{R}_{b}(t)-\mathbf{R}_{b'}(t))}\ket{\phi^{\mathbf{R}_{b}(t)}_{bj}},
\end{align}
the hopping matrix element is 
\begin{align}
&  \mathcal{H}^{\mathrm{TB}}_{bj,b'i}=\frac{t_{bj,b'i}}{2}\Big( e^{\frac{i}{\hbar}m\dot{\mathbf{R}}_{b'}(t) \cdot (\mathbf{R}_{b}(t)-\mathbf{R}_{b'}(t))}\nonumber\\&+e^{\frac{i}{\hbar}m\dot{\mathbf{R}}_{b}(t) \cdot (\mathbf{R}_{b}(t)-\mathbf{R}_{b'}(t))}\Big)\left(1-\delta_{bb'}\right).
\label{eq:Hamiltonian_hopping_VI}
\end{align}

\subsubsection{Lagrangian and Hamiltonian} 
The complete Ehrenfest Lagrangian for this tight-binding approximation is obtained from the LCAO Lagrangian (Eq. \eqref{eq:LCAO_lagrangian}) by applying the tight-binding approximation to the effective Hamiltonian and setting the overlap matrix diagonal $\mathcal{S}^{\mathrm{R,\dot{R}}}_{bj,b'i}=\delta_{b,b'}\delta_{ij}$
\begin{widetext}
    \begin{equation}
\begin{aligned}
 &\mathcal{L}^{\rm TB}=\sum_s \frac{(M_s+mN_{\rm el}^{\rm core})|\dot{\mathbf{R}}_s(t)|^2}{2}-\sum_{I=1}^{N_{\rm el}}\Bigg[
\sum_{\substack{bj,b'i}}\left(c^{I}_{bj}(t)\right)^* \mathcal{H}_{bj,b'i}^{\rm TB}c^I_{b'i}(t)
-\sum_{bj}\frac{i\hbar}{2}\Big(c^{I,*}_{bj}(t)\dot{c}^I_{bj}(t)-\dot{c}^{I,*}_{bj}(t)c^I_{bj}(t)\Big)\Bigg].
\label{eq:TB_lagrangian}
\end{aligned}
\end{equation}
The number of electrons in the core levels - those that are not included explicitly in the tight-binding model - is $N_{\rm el}^{\rm core}$. The core electrons' kinetic energy has a positive sign in the tight-binding Lagrangian since the electronic Hamiltonian enters in the Lagrangian with an overall negative sign (see Eq. \eqref{eq:lagrangian_general}), therefore reversing the negative sign of the nuclear velocity contribution to the on-site energy given in Eq. \eqref{eq:TB_onsite_VI}. The tight-binding Hamiltonian reads
\begin{equation}
  \mathcal{H}_{bj,b'i}^{\rm TB}=\frac{t_{bj,b'i}}{2}\Big( e^{\frac{i}{\hbar}m\dot{\mathbf{R}}_{b'}(t) \cdot (\mathbf{R}_{b}(t)-\mathbf{R}_{b'}(t))}
+e^{\frac{i}{\hbar}m\dot{\mathbf{R}}_{b}(t) \cdot (\mathbf{R}_{b}(t)-\mathbf{R}_{b'}(t))}\Big)
(1-\delta_{bb'})+\delta_{bb'}\delta_{ij}\left(\Delta_{b'i}-\frac{m}{2}|\dot{\mathbf{R}}_{b'}(t)|^2\right).
\label{eq:TB_hamiltonian}
\end{equation}
\end{widetext}
Compared with the adiabatic tight-binding Hamiltonian of Eq. \eqref{eq:TB_hamiltonian_adiabatic}, nuclear velocity-dependent phases appear on the hopping while the on-site energy is corrected by the kinetic energy of the electron comoving with the nucleus. The comparison is shown in Table \ref{tab:summary_eph}. 

The kinetic energy of the electrons comoving with the nucleus modifies the inertial mass in the nuclear dynamics, which also includes the electronic contribution. Refs. \cite{PhysRevX.7.031035, ponce2025searchelectronphononcontributiontotal} obtained the same result by keeping higher orders in the adiabatic approximation. Nevertheless, the electronic mass contribution is fully contained in the all-electron non-adiabatic vibrational response, as shown in Ref. \cite{PhysRevLett.136.196401}. The electronic mass term appears in our corrections to ensure the correspondence between the vibrational responses of the effective models and the all-electron system.

Since the hopping $t_{bj,b'i}$ depends on the positions of the moving nuclei, the electron-phonon coupling is evaluated by expanding $t_{bj,b'i}$ in terms of the displacement of the nuclei from the equilibrium position  \cite{PhysRevB.65.235412, PhysRevLett.93.185503,doi:10.1143/JPSJ.75.084713, PhysRevB.84.035433, PhysRevB.90.125414, Bistoni2019, PhysRevB.110.L201405, Fachin2025}.
The additional phases appearing in the hopping term are analogous to the Peierls phase \cite{PhysRevB.63.245101, PhysRevLett.91.196401}, introducing in the model an effective irrotational vector potential corresponding to an electric field. If all the nuclei were moving at the same velocity, the additional phases would describe an effective uniform electric field applied to the material in the vector potential gauge. 

\subsubsection{Linear response to nuclear displacement in the tight-binding approximation}
\label{sec:TB_derivatives}
Here, we compute the derivative of the tight-binding Hamiltonian in both the real and the reciprocal space that are used to assess the vibrational responses in the Section \ref{sec:vibrationa_responses}.  We remind that, in the context of linear response, the derivatives of the tight-binding Hamiltonian are computed with respect to the equilibrium configuration $\{\mathbf{R}_{b}(t=0),\dot{\mathbf{R}}_{b}=0\}$. 

The nuclear displacement derivatives of the hopping are usually determined by expanding it in terms of the nuclear displacement from equilibrium \cite{PhysRevB.65.235412, PhysRevLett.93.185503,doi:10.1143/JPSJ.75.084713, PhysRevB.84.035433, PhysRevB.90.125414, Bistoni2019, PhysRevB.110.L201405, Fachin2025, PhysRevB.53.15417}. Conversely, the derivatives with respect to the nuclear velocities are first reported in this work. In detail, the first-order nuclear velocity derivative is

\begin{equation}
\begin{aligned}
   \frac{\partial \mathcal{H}^{\rm TB}_{bj,b'i}}{\partial \dot{R}_{s\alpha}} =i\frac{m}{2\hbar} \left(R_{b\alpha}- R_{b'\alpha}\right)\mathcal{H}^{\rm TB}_{bj,b'i}  (\delta_{bs}+\delta_{b's}),
\end{aligned}
\label{eq:HTB_der_dotR}
\end{equation}
while the second-order derivatives are
\begin{align}
       &\frac{\partial \mathcal{H}^{\rm TB}_{bj,b'i}}{\partial R_{s\alpha}\partial \dot{R}_{s'\beta}}=i\frac{m}{2\hbar} \left(R_{b\beta}- R_{b'\beta}\right)\frac{\partial \mathcal{H}^{\rm TB}_{bj,b'i}}{\partial R_{s\alpha}}(\delta_{bs'}+\delta_{b's'}) \label{eq:2nd_derivative_mixed_real}\\
       &+i\frac{m}{2\hbar} \mathcal{H}^{\rm TB}_{bj,b'i}  \left( \delta_{bs}-\delta_{b's}\right) \left( \delta_{bs'}+\delta_{b's'}\right) \delta_{\alpha\beta}\nonumber
       \end{align}
       \begin{align}
      &  \frac{\partial \mathcal{H}^{\rm TB}_{bj,b'i}}{\partial \dot{R}_{s\alpha}\partial \dot{R}_{s'\beta}}=-\delta_{ss'}\Bigg(m\delta_{bb'}\delta_{bs}\delta_{\alpha\beta}\delta_{ij}+\frac{m^2}{2\hbar^2}\nonumber\\
      &\left(R_{b\alpha}- R_{b'\alpha}\right)\left(R_{b\beta}- R_{b'\beta}\right)\mathcal{H}^{\rm TB}_{bj,b'i}(\delta_{bs}+\delta_{b's})\Bigg). 
      \label{eq:2nd_derivative_real}
           \end{align}

The derivatives are also reported in Table \ref{tab:summary_eph}, summarising the main results of the paper and directly comparing the effect of the inclusion of nuclear velocity effects.

\textit{Reciprocal space expressions for zone centre phonons}--- In a crystalline system where each atomic site is identified by $\mathbf{R}_{s}=\mathbf{R}+\bm{\tau}_s$ with $\mathbf{R}$ the Bravais lattice vectors and $\bm{\tau}_s$ intra-cell vector, we define a basis of Bloch states from the localised atomic orbitals basis
\begin{equation}
    \ket{\mathbf{k},b,i}=\sum_{\mathbf{R}}e^{i\mathbf{k}\cdot \left( \mathbf{R}+\bm{\tau}_{b}\right)}\ket{\phi^{\dot{\mathbf{R}}_{b},\mathbf{R}_{b}}_{bi}}.
\end{equation}
By evaluating the real-space tight-binding Hamiltonian on the Bloch state basis, we obtain the reciprocal space Hamiltonian as
\begin{align}
    \mathcal{H}_{bj,b'i}^{\rm TB}({\mathbf{k}})=\sum_{\mathbf{R'}-\mathbf{R}}e^{i\mathbf{k}\cdot (\mathbf{R}'+\bm{\tau}_{b'}-\mathbf{R}-\bm{\tau}_{b})}\mathcal{H}^{\rm TB}_{bj,b'i}(\mathbf{R}'-\mathbf{R}).
\end{align}
Assuming that the nuclear displacement has a spatial modulation $e^{i\mathbf{q}\cdot \mathbf{R}_{s}}$, the tight-binding matrix elements in the reciprocal space are obtained as
\begin{equation}
    \begin{aligned}
     &\bra{\mathbf{k'},b,j}\frac{\partial \mathcal{H}^{\rm TB}}{\partial \dot{R}_{s\alpha}}e^{i\mathbf{q}\cdot \mathbf{R}_{s}}\ket{\mathbf{k},b',i}=\\
    &-\frac{m}{2\hbar}\Bigg(\delta_{bs}\frac{\partial \mathcal{H}^{\rm TB}_{bj,b'i}(\mathbf{k})}{\partial k_{\alpha}}+\frac{\partial \mathcal{H}^{\rm TB}_{bj,b'i}(\mathbf{k+q})}{\partial k_{\alpha}}\delta_{b's}\Bigg),
    \end{aligned}
    \label{eq:1st_order_derivative_kspace_finiteq}
\end{equation}

that reduces for a zone centre phonon to
\begin{align}
  \frac{\partial \mathcal{H}^{\rm TB}_{bj,b'i}}{\partial \dot{R}_{s'\alpha}}(\mathbf{k})=-\frac{m}{2\hbar}\frac{\partial \mathcal{H}^{\rm TB}_{bj,b'i}(\mathbf{k})}{\partial k_{\alpha}}(\delta_{bs'}+\delta_{b's'}).
    \label{eq:1st_order_derivative_kspace}
\end{align}
The second order derivative for a zone-centre phonon displacement is 

    \begin{align}
        \frac{\partial^2 \mathcal{H}^{\rm TB}_{bj,b'i}(\mathbf{k})}{\partial R_{s\alpha}\partial \dot{R}_{s'\beta}}&= -\frac{m}{2\hbar}\frac{\partial^2 \mathcal{H}^{\rm TB}_{bj,b'i}(\mathbf{k})}{\partial R_{s\alpha}\partial k_{\beta}}(\delta_{bs'}+\delta_{b's'})\label{eq:2nd_derivative_mixed_reciprocal}\\
 +i\frac{m}{2\hbar}&\mathcal{H}^{\rm TB}_{bj,b'i}  (\mathbf{k})\left( \delta_{bs}-\delta_{b's}\right) \left( \delta_{bs'}+\delta_{b's'}\right) \delta_{\alpha\beta},\nonumber
 \end{align}
 \begin{align}
        \frac{\partial^2 \mathcal{H}^{\rm TB}_{bj,b'i}(\mathbf{k})}{\partial \dot{R}_{s\alpha}\partial \dot{R}_{s'\beta}}&=\Bigg(\frac{m^2}{2\hbar^2}\frac{\partial \mathcal{H}^{\rm TB}_{bj,b'i}(\mathbf{k})}{\partial k_{\alpha}\partial k_{\beta}}(\delta_{bs}+\delta_{b's})\label{eq:2nd_derivative_reciprocal}\\&-m\delta_{bb'}\delta_{bs}\delta_{\alpha\beta}\delta_{ij}\Bigg)\delta_{ss'}.\nonumber
    \end{align}
\subsubsection{Linear response to an electric field in the tight-binding approximation}
\label{sec:TB_electric field}
We consider the coupling to the electric field $\bm{\mathcal{E}}(t)$ in the scalar potential gauge, where, assuming that the position operator is diagonal according to Eq. \eqref{eq:diagonal_pos_operator}, the coupling Hamiltonian in the real space reads 
\begin{equation}
     \left(\mathcal{H}^{\rm TB}_{\rm em}\right)_{bj,b'i}=e(\mathbf{R}+\bm{\tau}_s) \cdot \bm{\mathcal{E}}(t) \delta_{ss'}\delta_{ij}.
     \label{eq:electric_field_hamiltonian}
\end{equation}
Therefore, the electric field derivative of the tight-binding Hamiltonian in the real space is straightforward, corresponding to a diagonal contribution in the atomic orbital basis proportional to the position of the nucleus,  
\begin{equation}
   \left( \frac{\partial \mathcal{H}^{\rm TB}_{\rm em}}{\partial \mathcal{E}_{\alpha}}\right)_{bj,b'i}=e\left(R_{\alpha}+(\tau_s)_{\alpha}\right)\delta_{ss'}\delta_{ij}.
   \label{eq:electric_field_derivative}
\end{equation}
Transforming to the reciprocal space, the electric field derivative of the Hamiltonian can be related to the band velocity  \cite{Vanderbilt2018, 10.21468/SciPostPhys.12.2.070, fachin2026lattice}
\begin{equation}
\Bigg[\mathcal{H}^{\rm TB}(\mathbf{k}),  \frac{\partial \mathcal{H}^{\rm TB}_{\rm em}}{\partial \mathcal{E}_{\alpha}}(\mathbf{k})\Bigg]=i\frac{\partial  \mathcal{H}^{\rm TB}(\mathbf{k})}{\partial \mathbf{k}} .
\end{equation}

\subsection{Velocity-including Dirac Hamiltonian}
\label{sec:kp_models}
In this Section, we show how to introduce the nuclear velocity-dependence to linear order in the low-energy Dirac Hamiltonians, describing the low-energy physics around the valley points ( $\rm \mathbf{K}$ and $\rm \mathbf{K}'$) of crystals with honeycomb lattice and a diatomic basis, such as gapped graphene and the topologically non-trivial Haldane and Kane-Mele models \cite{Bernevig2013}.  They are obtained from the expansion of the tight-binding reciprocal space Hamiltonian around the two valleys ( $\rm \mathbf{K}$ and $\rm \mathbf{K}'$) - described by the valley index $\eta=\pm1$ respectively - in terms of  $\mathbf{p}=\mathbf{k}-\mathbf{K}(\mathbf{K}')$. The electronic Hamiltonian is described in terms of the Pauli matrices
\begin{equation}
    \begin{aligned}
    H_\eta^{\rm D,el}(\mathbf{p})&=\hbar v_F \Bigg(\eta p_x\sigma^\mathrm{P}_x+p_y\sigma^\mathrm{P}_y\Bigg)+\Delta(\eta)\sigma^\mathrm{P}_z,
    \label{eq:low_energy}
\end{aligned}
\end{equation}
where $v_F$ is the Fermi velocity; $\Delta(\eta)=\frac{\Delta}{2}$ for gapped graphene ($\Delta=0$ for graphene), $\Delta(\eta)=\frac{\Delta}{2}-\eta 3\sqrt{3}t_2$ for the Haldane model with $t_2$ imaginary second-nearest neighbour hopping; the Pauli matrices $\sigma^\mathrm{P}$ describe the sublattice degree of freedom with the two sites labelled as A and B respectively. The electron-phonon coupling, expressed in terms of the displacement of the nuclei from the equilibrium position $\mathbf{u}_{s\alpha}(t)=\mathbf{R}_{s\alpha}(t)-\mathbf{R}_{s\alpha}(t=0)$, enters as a gauge field around the valleys \cite{Bistoni2019,  PhysRevB.110.L201405, Fachin2025, PhysRevB.76.045430, PhysRevB.90.125414, PhysRevB.65.235412, PhysRevLett.93.185503}
\begin{equation}
    \begin{aligned}
    H_\eta^{\rm D,e-ph}=&\hbar v_F \xi_{\rm e-ph} \Bigg(-\left(u_{A,y}(t)-u_{B,y}(t)\right)\sigma^\mathrm{P}_x\\&
    +\eta \left(u_{A,x}(t)-u_{B,x}(t)\right)\sigma^\mathrm{P}_y\Bigg).
    \label{eq:low_energy_eph}
\end{aligned}
\end{equation}
where $\xi_{\rm e-ph}$ is the electron-phonon coupling constant. Usually the Dirac Hamiltonian with the electron-phonon coupling are obtained as $H^D_\eta=H_\eta^{\rm D,el}(\mathbf{p})+H_\eta^{\rm D, e-ph}$\cite{PhysRevB.110.L201405}. In addition to the electron-phonon coupling originating from the nuclear displacement, the linear expansion in terms of the nuclear velocity of the tight-binding Hamiltonian, using Eq. \eqref{eq:1st_order_derivative_kspace}, yields 
\begin{equation}
    \begin{aligned}
    H_\eta^{\rm D,e-ph,\dot{R}}=&-\frac{1}{2}m v_F \Bigg( \eta \left(\dot{u}_{A,x}(t)+\dot{u}_{B,x}(t)\right)\sigma^\mathrm{P}_x\\&+\left(\dot{u}_{A,y}(t)+\dot{u}_{B,y}(t)\right)\sigma^\mathrm{P}_y\Bigg).
    \label{eq:low_energy_eph_dotR}
\end{aligned}
\end{equation}
The nuclear-velocity derivative of the low-energy Hamiltonian is directly obtained from the above Hamiltonian as
\begin{equation}
    \frac{\partial H^{\mathrm{D}}(\mathbf{p})}{\partial   \dot{R}_{s\alpha}}=-\frac{1}{2} m v_{F}\sigma^\mathrm{P}_{\alpha}\left(\eta\delta_{\alpha x}+\delta_{\alpha y} \right).
    \label{eq:Dirac_vel_derivative}
\end{equation}
 The effect of the nuclear velocities on the Dirac low energy-Hamiltonians is summarised in Table \ref{tab:summary_eph}.
In low-energy models, the correction to the electron-phonon coupling has a clear and simple expression, being proportional to the band velocity.  In other words, a velocity vertex with a proper prefactor has to be added to the usual electron-phonon coupling vertex. 
This result can be straightforwardly generalised to other continuous low energy models. 

\section{Vibrational responses}
\label{sec:vibrationa_responses}
In this Section, we study the corrections to the frequency-dependent vibrational responses, introduced by the nuclear velocity dependence in the tight-binding and Dirac low-energy Hamiltonian, focusing on the force constant matrix and the Born effective charges. To this aim, we derive explicit expressions for the first- and second-order nuclear-displacement derivatives of the tight-binding and Dirac low-energy Hamiltonians, enabling evaluation of forces and vibrational responses.  

\subsection{Born effective charges}
The Born effective charges quantify the coupling between light and lattice excitation in optical spectra
\cite{PhysRevLett.68.3603,PhysRevB.58.6224}. When the lattice excitations are resonant with the electronic ones, the frequency-dependence of the Born effective charges has to be accounted for \cite{PhysRevB.86.115439,Bistoni2019, PhysRevB.110.L201405, Fachin2025, PhysRevB.103.134304, PhysRevB.106.L180303}. Born effective charges are defined as the variation of the electronic polarization due to the time-dependent displacement of an individual ion $s$  or, alternatively, as the variation of the force acting on the ion $s$ due to an external electric field
\begin{equation}
    e\mathcal{Z}^*_{\alpha,s\beta}(t-t')=V_{\rm c}\frac{\delta P_{\alpha}(t)}{\delta R_{s\beta}(t')}
    \label{eq:BEC_definition}
\end{equation}
where $\delta$ indicates the functional derivative and $V_{\rm c}$ is the unit cell volume. The Born effective charges are the sum of a rigid ionic contribution $\mathcal{Z}^{\rm{ion}}$, accounting for the displacement of the electric charge on the nucleus, and an electronic contribution $\mathcal{Z}^{\rm el}$ related to the polarization induced in the distribution of valence electrons by the atomic displacement:
\begin{align}
\mathcal{Z}^*_{\alpha,s\beta}=\mathcal{Z}^{\rm{ion}}_{s}\delta_{\alpha\beta}+\mathcal{Z}^{\rm{el}}_{\alpha,s\beta}.
\label{eq:BEC_definition_ionic_electronic}
\end{align}
Expanding the polarization in terms of the nuclear position and velocity---acceleration does not contribute since it does not appear in the tight-binding Hamiltonian in the diagonal position operator assumption---we obtain
\begin{align}
    \delta P_\alpha(t)=\int_{0^+}^{t} dt' \Bigg(\frac{\partial P_{\alpha}}{\partial R_{s\beta}}(t-t') \delta R_{s\beta}(t')\nonumber\\
   + \frac{\partial P_{\alpha}}{\partial \dot{R}_{s\beta}}(t-t')\delta \dot{R}_{s\beta}(t')\Bigg).
\end{align}
Integrating by parts the variation of the nuclear velocity, we obtain
\begin{align}
     \delta P_\alpha(t)=\int_{0^+}^{t} dt' \Bigg[\frac{\partial P_{\alpha}}{\partial R_{s\beta}}(t-t') \nonumber\\-\frac{d}{dt'} \left(\frac{\partial P_{\alpha}}{\partial \dot{R}_{s\beta}}(t-t')\right)\Bigg] \delta R_{s\beta} (t').
     \label{eq:pol_expansion_time}
\end{align}
Therefore, the time-dependent Born effective charges are 
\begin{align}
\frac{e}{V_{\rm c}}\mathcal{Z}^*_{\alpha,s\beta}(t-t')=\frac{\partial P_{\alpha}}{\partial R_{s\beta}}(t-t') \nonumber\\
-\frac{d}{dt'} \left(\frac{\partial P_{\alpha}}{\partial \dot{R}_{s\beta}}(t-t')\right).
    \label{eq:td-BEC}
\end{align}
The Fourier transform is defined as $\displaystyle f(\omega+i\zeta)=\int_{-\infty}^{+\infty} f(t) e^{i(\omega+i\zeta) t}$. Because of the causality $\displaystyle f(t)=0$ for $\displaystyle t<0$, therefore the transform is well-defined for $\displaystyle \zeta>0$. In the following, we assess the physical observables  in the limit of $\displaystyle \zeta\to 0^+$, expressing them (as a shorthand notation) just as a function of $\displaystyle \omega$. 

The Fourier transform of Eq.\eqref{eq:td-BEC} gives the frequency-dependent Born effective charges 
\begin{align}
    \frac{e}{V_c}\mathcal{Z}^*_{\alpha,s\beta}(\omega)=  \frac{\partial P_{\alpha}}{\partial R_{s\beta}}(\omega) -i\omega \frac{\partial P_{\alpha}}{\partial \dot{R}_{s\beta}}(\omega).
\end{align}
In the following, we describe the interaction of the system with an external electric field $\mathcal{E}_{\alpha}$, acting in the Cartesian direction $\alpha$, in the scalar potential gauge, according to the Eqs. \eqref{eq:electric_field_hamiltonian} and \eqref{eq:electric_field_derivative}. We define the following matrix elements of the tight-binding Hamiltonian derivatives of the electric field, nuclear displacement and velocity as
\begin{equation}
M_{\alpha}^{IJ}=\sum_{bj,b'i}\left(c^{J}_{bj}\right)^*\frac{\partial \mathcal{H}_{bj,b'i}^{\rm TB}}{\partial \mathcal{E}_{\alpha}}c^I_{b'i}. 
\end{equation}
\begin{equation}
g_{s\beta}^{IJ}=\sum_{bj,b'i}\left(c^{J}_{bj}\right)^*\frac{\partial \mathcal{H}_{bj,b'i}^{\rm TB}}{\partial R_{s\beta}}c^I_{b'i},
\end{equation}
\begin{equation}
    f_{s\beta}^{IJ}=\sum_{bj,b'i}\left(c^{J}_{bj}\right)^*\frac{\partial \mathcal{H}_{bj,b'i}^{\rm TB}}{\partial \dot{R}_{s\beta}}c^I_{b'i},
\end{equation}
The above vertices are static in a tight-binding model since we are not considering any self-consistent dressing of the interaction in the response.

The electronic contribution to the frequency-dependent Born effective charges in linear response theory reads \cite{Bistoni2019, Fachin2025}
\begin{equation}
\begin{aligned}
&e\mathcal{Z}^{\mathrm{el}}_{\alpha,s\beta}(\omega)=-\rho_s\delta_{\alpha\beta}\\
  &-2\sum_{I,J} \frac{f_{ I}-f_{ J}}{ \epsilon_{I}-\epsilon_{J}+\hbar \omega}M_{\alpha}^{IJ}\left(g_{s\beta}^{JI}-i\omega f_{s\beta}^{JI} \right)
\end{aligned}
    \label{eq:cariche_efficaci}
\end{equation}
where $\epsilon_{I}$ are the tight-binding energies corresponding to the coefficients $\mathbf{c}^I$  with Fermi-Dirac statistical weight $f_I$;  $\rho_s$ is the electronic charge density on the $s$ site defined as the sum of squared moduli of the tight-binding coefficients of the occupied states $\rho_s=\sum_{J,j}f_J |c^J_{sj}|^2$. Its summation over the sites gives the total number of electrons in the system $\sum_s \rho_s=N_{\rm el}$.
The above expression is equivalent to the time-dependent density functional perturbation theory (DFPT) result  \cite{PhysRevB.82.165111, PhysRevB.103.134304}, when extended for moving nuclei as done in Ref.  \cite{PhysRevLett.136.196401}; this is seen by identifying the wavefunction $\mathbf{c}^I\to \ket{\psi_i}$ and the Hamiltonian operator $\mathcal{H}^{\rm TB} \to H$.  
Furthermore, we remark that the self-consistency has to be properly accounted for in a DFPT approach \cite{RevModPhys.73.515,RevModPhys.89.015003, PhysRevB.82.165111, PhysRevB.111.075137}, while in the tight-binding model it does not.

From the above results we can say that, within the velocity-including atomic orbital approach, the tight-binding electron-phonon coupling has an additional contribution compared to the adiabatic case, changing as 
\begin{equation}
 \frac{\partial \mathcal{H}^{\rm TB}}{\partial R_{s\beta}}\to     \frac{\partial \mathcal{H}^{\rm TB}}{\partial R_{s\beta}} -i\omega\frac{\partial \mathcal{H}^{\rm TB}}{\partial \dot{R}_{s\beta}},
 \label{eq:eph_vertex_velocity_correction_TB}
\end{equation}
that, using Eq. \eqref{eq:HTB_der_dotR}, explicitly is 
\begin{equation}
\begin{aligned}
   \frac{\partial \mathcal{H}^{\rm TB}_{bj,b'i}}{\partial R_{s\alpha}} \to   \frac{\partial \mathcal{H}^{\rm TB}_{bj,b'i}}{\partial R_{s\alpha}} +\omega\frac{m}{2\hbar} \left(R_{b\alpha}- R_{b'\alpha}\right)\mathcal{H}^{\rm TB}_{bj,b'i}  (\delta_{bs}+\delta_{b's})
\end{aligned}
\label{eq:eph_1st_TB_real_space}
\end{equation}
and in reciprocal space
\begin{align}
  \frac{\partial \mathcal{H}^{\rm TB}_{bj,b'i}(\mathbf{k})}{\partial R_{s'\alpha}}  \to\frac{\partial \mathcal{H}^{\rm TB}_{bj,b'i}(\mathbf{k})}{\partial R_{s'\alpha}}\nonumber\\
  +i\omega\frac{m}{2\hbar}\frac{\partial \mathcal{H}^{\rm TB}_{bj,b'i}(\mathbf{k})}{\partial k_{\alpha}}(\delta_{bs'}+\delta_{b's'}).
    \label{eq:1st_order_derivative_kspace_sub}
\end{align}
In the case of a Dirac Hamiltonian for systems like (gapped) graphene or the Haldane model - described in detail in Section \ref{sec:kp_models}-  using Eq. \eqref{eq:Dirac_vel_derivative}, the electron-phonon coupling becomes 
\begin{align}
    \frac{\partial H^{\mathrm{D}}(\mathbf{p})}{\partial R_{s\alpha}}&\to  \frac{\partial H^{\mathrm{D}}(\mathbf{p})}{\partial R_{s\alpha}}+\frac{1}{2}i\omega m v_{F}\sigma^\mathrm{P}_{\alpha}\left(\eta\delta_{\alpha x}+\delta_{\alpha y} \right),
    \label{eq:Dirac_1st_derivative_eph}
\end{align}
where $\eta=\pm1$ is the valley index. 
 
\subsection{Force constant matrix}
The time-dependent force constant matrix, determining the phonon frequency and its lifetime, is defined as the functional derivative of the time-dependent force acting on the $s'$ nucleus due to the time-dependent displacement of the nucleus $s$ \cite{PhysRevB.82.165111, PhysRevLett.136.196401}:
\begin{equation}
    \mathcal{C}_{s\alpha,s'\beta}(t-t')=-\frac{\delta F_{s'\beta}(t)}{\delta R_{s\alpha}(t')}.
    \label{eq:def_force_constant_matrix}
\end{equation}
In the frequency space, the force constant matrix can be conveniently split into the zero-frequency (usually called `adiabatic') contribution and the frequency-dependent self-energy, accounting for the dynamical effects
$\mathcal{C}_{s\alpha,s'\beta}(\omega)=\mathcal{C}_{s\alpha,s'\beta}(\omega=0)+\Pi_{{s\alpha},{s'\beta}}(\omega)$ , where the zero-frequency contribution is the sum of the bare force-constant matrix and of the static phonon-self energy $\mathcal{C}_{s\alpha,s'\beta}(\omega=0)=\mathcal{C}^b_{{s\alpha},{s'\beta}}+\Pi_{{s\alpha},{s'\beta}}(\omega=0)$\cite{PhysRevB.82.165111, RevModPhys.89.015003, PhysRevB.111.075137, RevModPhys.73.515} . Only the frequency-dependent self-energy $\Pi_{{s\alpha},{s'\beta}}(\omega)$ includes nuclear velocity-dependent correcting terms, absent in the static response. 
Focusing on the optical response, the zone centre phonon frequency of the mode $\nu$, $\omega_\nu$, is determined by solving the self-consistent eigenvalue equation with the hermitian part of the force constant matrix
\begin{equation}
    \mathrm{Det}\Bigg| \frac{\mathcal{C}_{s\alpha,s'\beta}(\omega_\nu)+\mathcal{C}^*_{{s'\beta},{s\alpha}}(\omega_\nu)}{2\sqrt{M_sM_{s'}}} -\omega_\nu^2\delta_{\alpha\beta}\delta_{ss'}\Bigg| =0.
    \label{eq:ph_freq}
\end{equation}
The phonon linewidth of the mode $\nu$, $\gamma_{\nu}$, is instead determined by the anti-Hermitian part of the force constant matrix \cite{PhysRevB.82.165111}
\begin{equation}
    \gamma_{\nu}=\frac{2}{\omega_{\nu}}\sum_{ss'}e_{s\alpha,\nu}\left(\frac{\mathcal{C}_{s\alpha,s'\beta}(\omega_\nu)-\mathcal{C}^*_{{s'\beta},{s\alpha}}(\omega_\nu)}{2i\sqrt{M_sM_{s'}}}\right)e_{s'\beta,\nu},
        \label{eq:ph_lifetime}
\end{equation}
where $e_{s\alpha,\nu}$ are the eigenvectors corresponding to the phonon frequency $\omega_{\nu}$. 

As for the Born effective charges, we express the force-constant matrix in the frequency space in terms of the action defined in Section \ref{sec:II}
    \begin{align}
     &M_{s\alpha,s'\beta}\omega^2-\mathcal{C}_{s\alpha,s'\beta}(\omega)=\frac{\partial^2 \mathcal{L}^{\rm TB}}{\partial R_{s\alpha}\partial R_{s'\beta}}
     \label{eq:force_constant_matrix_lagrangian}\\
     &-i\omega \frac{\partial^2 \mathcal{L}^{\rm TB}}{\partial R_{s\alpha}\partial \dot{R}_{s'\beta}}+i\omega \frac{\partial^2 \mathcal{L}^{\rm TB}}{\partial \dot{R}_{s\alpha}\partial R_{s'\beta}}+\omega^2\frac{\partial^2 \mathcal{L}^{\rm TB}}{\partial \dot{R}_{s\alpha}\partial \dot{R}_{s'\beta}}\nonumber
\end{align}
where $M_{s\alpha,s'\beta}=M_s\delta_{ss'}\delta_{\alpha\beta}$. We remark that, in the absence of nuclear velocity effects, only the first term is present. From Eq. \eqref{eq:force_constant_matrix_lagrangian}, we obtain the linear response expressions for those contributions as
\begin{widetext}

\begin{equation}
\begin{aligned}
     \mathcal{C}_{s\alpha,s'\beta}(\omega) =&2\sum_{J} f_{J}\sum_{bi,b'j}\left(c^{J}_{bi}\right)^*\Bigg[\frac{\partial^2 \mathcal{H}_{bi,b'j}^{\rm TB}}{\partial R_{s,\alpha}\partial R_{s',\beta}}-i\omega\left(\frac{\partial^2 \mathcal{H}_{bi,b'j}^{\rm TB}}{\partial R_{s,\alpha}\partial \dot{R}_{s',\beta}}-\frac{\partial^2 \mathcal{H}_{bi,b'j}^{\rm TB}}{\partial \dot{R}_{s,\alpha}\partial R_{s',\beta}}\right)+\omega^2\frac{\partial^2 \mathcal{H}_{bi,b'j}^{\rm TB}}{\partial \dot{R}_{s,\alpha}\partial \dot{R}_{s',\beta}}\Bigg]c^{J}_{b'j}\\&+2\sum_{I,J} \frac{f_{I}-f_{J}}{\epsilon_{I}-\epsilon_{J}+\hbar\omega}\left(g_{s\alpha}^{IJ}+i\omega f_{s\alpha}^{IJ} \right)\left(g_{s'\beta}^{JI}-i\omega f_{s'\beta}^{JI} \right).
\end{aligned}
\label{eq:force_constant_matrix}
\end{equation}

\end{widetext}
As for Born effective charges, the above expressions are formally equivalent to the DFPT expression with the proper substitutions discussed above. Therefore, the second-order nuclear displacement derivative of the Hamiltonian changes, for the tight-binding, as
\begin{equation}
\begin{aligned}
     & \frac{\partial^2 \mathcal{H}^{\rm TB}}{\partial R_{s,\alpha}\partial R_{s',\beta}} \to   \frac{\partial^2 \mathcal{H}^{\rm TB}}{\partial R_{s,\alpha}\partial R_{s',\beta}}-i\omega\Bigg(\frac{\partial^2 \mathcal{H}^{\rm TB}}{\partial R_{s,\alpha}\partial \dot{R}_{s',\beta}}\\&-\frac{\partial^2 \mathcal{H}^{\rm TB}}{\partial \dot{R}_{s,\alpha}\partial R_{s',\beta}}\Bigg)+\delta_{ss'}\omega^2\frac{\partial^2 \mathcal{H}^{\rm TB}}{\partial \dot{R}_{s,\alpha}\partial \dot{R}_{s',\beta}}
\end{aligned}
\label{eq:eph_2nd_TB_general}
\end{equation}
that explicitly, using Eqs.\eqref{eq:2nd_derivative_mixed_real}, \eqref{eq:2nd_derivative_real}, is

    \begin{align}
  &\frac{\partial^2 \mathcal{H}^{\rm TB}_{bj,b'i}}{\partial R_{s\alpha}\partial R_{s'\beta}}\to \frac{\partial^2 \mathcal{H}^{\rm TB}_{bj,b'i}}{\partial R_{s\alpha}\partial R_{s'\beta}}+
 \nonumber   \\
    &+\omega\frac{m}{2\hbar} \left(R_{b\beta}- R_{b'\beta}\right)\frac{\partial \mathcal{H}^{\rm TB}_{bj,b'i}}{\partial R_{s\alpha}}(\delta_{bs'}+\delta_{b's'})\nonumber \\
 &-\omega\frac{m}{2\hbar} \left(R_{b
\alpha}- R_{b'\alpha}\right)\frac{\partial \mathcal{H}^{\rm TB}_{bj,b'i}}{\partial R_{s'\beta}}(\delta_{bs}+\delta_{b's})
\label{eq:eph_2nd_TB_real_space}\\&
+\omega\frac{m}{\hbar}\mathcal{H}^{\rm TB}_{bj,b'i}   \left( \delta_{bs}\delta_{b's'}-\delta_{bs'}\delta_{b's}\right) \delta_{\alpha\beta}\nonumber\\&
-\omega^2\delta_{ss'}\Bigg(m\delta_{bb'}\delta_{bs}\delta_{\alpha\beta}\delta_{ij}+\frac{m^2}{2\hbar^2}\left(R_{b\alpha}- R_{b'\alpha}\right)\nonumber \\
&\times\left(R_{b\beta}- R_{b'\beta}\right)\mathcal{H}^{\rm TB}_{bj,b'i}(\delta_{bs}+\delta_{b's})\Bigg).\nonumber 
\end{align}
In the reciprocal space for zone centre phonons, using Eqs.\eqref{eq:2nd_derivative_mixed_reciprocal} and \eqref{eq:2nd_derivative_reciprocal}, 
\begin{align}
        &\frac{\partial^2 \mathcal{H}^{\rm TB}_{bj,b'i}(\mathbf{k})}{\partial R_{s\alpha}\partial R_{s'\beta}} \to  \frac{\partial^2 \mathcal{H}^{\rm TB}_{bj,b'i}(\mathbf{k})}{\partial R_{s\alpha}\partial R_{s'\beta}}\nonumber \\
        &-i\omega\frac{m}{2\hbar}\Bigg(\frac{\partial^2 \mathcal{H}^{\rm TB}_{bj,b'i}(\mathbf{k})}{\partial k_{\alpha}\partial R_{s'\beta}}(\delta_{bs}+\delta_{b's})\label{eq:reciprocal_space_diamagnetic}\\
    &-\frac{\partial^2 \mathcal{H}^{\rm TB}_{bj,b'i}(\mathbf{k})}{\partial k_{\beta}\partial R_{s\alpha}}(\delta_{bs'}+\delta_{b's'})\Bigg)\nonumber\\
   &+\omega\frac{m}{\hbar}\mathcal{H}^{\rm TB}_{bj,b'i}  (\mathbf{k}) \left( \delta_{bs}\delta_{b's'}-\delta_{bs'}\delta_{b's}\right) \delta_{\alpha\beta}\nonumber\\&+\omega^2\delta_{ss'} \left(\frac{m^2}{2\hbar^2}\frac{\partial \mathcal{H}^{\rm TB}_{bj,b'i}(\mathbf{k})}{\partial k_{\alpha}\partial k_{\beta}}(\delta_{bs}+\delta_{b's})-m\delta_{bb'}\delta_{bs}\delta_{\alpha\beta}\delta_{ij}\right)\nonumber 
   .
    \end{align}

\subsection{Sum rules on the dynamical matrix and Born effective charges for the tight-binding and Dirac Hamiltonians}\label{sec:sum_rules}
If the nuclear-velocity corrections are not kept into account, the sum rules for the frequency-dependent Born effective charges and dynamical matrix vanish in tight-binding and continuous low-energy models \cite{Bistoni2019, Fachin2025}. We show that the corrections introduced in this work restore the all-electron sum rules in these models. 

The optical conductivity is defined $\sigma_{\alpha \beta}(\omega)$ as the electric field derivative of the current density $\mathbf{J}$ 
\begin{equation}
    \sigma_{\alpha \beta}(\omega)=\frac{\partial J_{\alpha}}{\partial \mathcal{E}_{\beta}}(\omega), 
\end{equation}
which yields the tight-binding  linear response expression 
\begin{equation}
\begin{aligned}
  &\sigma^{\mathrm{el}}_{\alpha\beta}(\omega)=-\frac{2i\omega}{V_c}\sum_{I,J} \frac{f_{ I}-f_{ J}}{ \epsilon_{I}-\epsilon_{J}+\hbar \omega}M_{\alpha}^{IJ}M_{\beta}^{JI}.
\end{aligned}
    \label{eq:optical_conductivity}
\end{equation}
where $V_c$ is the unit cell volume (or area in 2D). The summation over the entire lattice of the nuclear velocity derivative of the tight-binding Hamiltonian (Eq. \eqref{eq:eph_1st_TB_real_space}) corresponds to the application of an electric field in the vector potential gauge
\begin{align}
    &-i\omega\sum_{s} \frac{\partial \mathcal{H}_{bj,b'i}^{\rm TB}}{\partial \dot{\mathbf{R}}_{s}}(\omega)=\omega \frac{m}{\hbar}\left(\mathbf{R}_{b}-\mathbf{R}_{b'}\right)\mathcal{H}^{\rm TB}_{bj,b'i}\nonumber\\&=-i\omega\frac{m}{e} \frac{\partial \mathcal{H}_{bj,b'i}^{\rm TB}}{\partial \mathbf{A}(\omega)}, 
\end{align}
where $\mathbf{A}(\omega)$ is a vector potential describing an electric field at frequency $\omega$, entering in the Hamiltonian via the Peierls substitution \cite{PhysRevB.51.4940}.
Then, the assumption of a diagonal position operator, defined in Eq. \eqref{eq:diagonal_pos_operator}, is compatible only with the hypothesis that the atomic orbitals are delta-like functions centred at the atomic sites. Accordingly, the electron-phonon coupling can depend only on the distance between the sites. Therefore, if all the sites of the lattice are displaced by the same amount, the hopping does not change. This implies that, under such approximations for tight-binding and continuous models,
\begin{equation}
    \sum_s\frac{\partial P_{\alpha}}{\partial R_{s\beta}}(\omega)=0.
    \label{eq:zero_sum_rule_TB}
\end{equation}

Consequently, the summation over the entire lattice of the Born effective charges gives
\begin{align}
   &\sum_s \frac{e\mathcal{Z}^*_{\alpha,s\beta}(\omega)}{V_{\rm c}}=\sum_s\left(\frac{\partial P_{\alpha}}{\partial R_{s\beta}}(\omega)-i\omega\frac{\partial P_{\alpha}}{\partial \dot{R}_{s\beta}}(\omega)\right)\nonumber\\
   &=-i\omega\frac{m}{e}\frac{\partial P_{\alpha}}{\partial A_{\beta}}(\omega)=-i\omega \frac{m}{e}\sigma_{\alpha \beta}(\omega),
   \label{eq:BEC_sum_rule}
\end{align}
where we used Eq. \eqref{eq:zero_sum_rule_TB} and $\displaystyle \frac{\partial P_{\alpha}}{\partial A_{\beta}}(\omega)=\frac{\partial J_{\alpha}}{\partial E_{\beta}}(\omega)$. Eq.  \eqref{eq:BEC_sum_rule} restores the sum rule for the all-electron Hamiltonian derived in the \textit{ab initio} framework \cite{PhysRevLett.128.095901, Marchese2024,PhysRevLett.129.185902, PhysRevB.107.094308, PhysRevB.110.094306, PhysRevLett.136.196401}. Summarising, in a tight-binding framework, the sum rule for the frequency-dependent Born effective charges is completely determined by the nuclear velocity correction.  

Following the same procedure, using that the Born effective charges are alternatively defined as the functional derivative of a force with respect to a time-dependent electric field
\begin{equation}
    e\mathcal{Z}^*_{\alpha,s\beta}(\omega)=\frac{\partial F_{s\beta}}{\partial \mathcal{E}_{\alpha}}(-\omega),
    \label{eq:BEC_definition_force}
\end{equation}
and the force constant matrix, defined in Eq. \eqref{eq:def_force_constant_matrix}, is expressed as 
\begin{equation}
    \mathcal{C}_{s\alpha,s'\beta}(\omega)=-\frac{\partial F_{s'\beta}}{\partial R_{s\alpha}}(\omega)+i\omega\frac{\partial F_{s'\beta}}{\partial \dot{R}_{s\alpha}}(\omega),
    \label{eq:def_force_constant_matrix_force_der}
\end{equation}
we obtain the tight-binding sum rules for the frequency-dependent force constant matrix that coincide with the all-electron one \cite{PhysRevLett.136.196401} 
\begin{equation}
    \begin{aligned}
        &\sum_s \mathcal{C}_{s\alpha,s'\beta}(\omega)=m\omega^2 \mathcal{Z}^{\rm el}_{\alpha,s'\beta}(-\omega)
        ,\\
        &\sum_{s'} \mathcal{C}_{s\alpha,s'\beta}(\omega)=m\omega^2 \mathcal{Z}^{ \rm el}_{\beta,s\alpha}(\omega).
        \label{eq:sum_rules_force_constant_matrix}
\end{aligned}
\end{equation}
The summation over both indices is obtained by using the Born effective charges sum rule for the summation over the second index 
  \begin{equation}
    \sum_{ss'} \mathcal{C}_{s\alpha,s'\beta}(\omega)=-\omega^2m N_{\rm el}
    -i\omega^3\frac{m^2}{e^2}V_{\rm c} \sigma_{\alpha\beta}(\omega)
\end{equation}
where $N_{\rm el}=\sum_s \rho_s$ is the total electric charge of the electrons in the system. The sum rule entails the presence of electronic inertia in the nuclear problem, via the adiabatic renormalisation of the mass, a result also achieved by keeping higher orders of the adiabatic approximation in the Born-Oppenheimer framework \cite{PhysRevX.7.031035,ponce2025searchelectronphononcontributiontotal}, and the non-adiabatic correction accounting for the conducting electrons that are left behind in the lattice translation \cite{PhysRevLett.128.095901, PhysRevLett.136.196401}.  
\\
The sum rules with the nuclear displacement derivatives of the Dirac Hamiltonian presented in Section \ref{sec:kp_models} yields the same result with analogous calculations.
\\
The restoration of the all-electron sum rules on frequency-dependent vibrational responses demonstrates the necessity to include these corrections in tight-binding models and Dirac Hamiltonians. While the corrections to the vibrational sum-rules in \textit{ab initio} calculations are small because the non-locality of the potential is limited to a short range around the nuclei \cite{PhysRevLett.136.196401}, conversely, the complete non-locality of tight-binding and low-energy models implies that the sum rule is fully accounted for by the nonadiabatic corrections, being otherwise exactly zero. These corrections are, therefore, needed in metallic systems or when the lattice excitations resonate with the electronic ones.

\begin{widetext}

\begin{table}[h!]
    \centering
    
    \begin{center}
    WITHOUT NUCLEAR VELOCITY CORRECTIONS
    \end{center}

    \renewcommand{\arraystretch}{2}
   \begin{tabular}{|c||c|}
    \cline{1-2}
    All-electron & Tight-binding \\    
      \cline{1-2} \rule[-0.6cm]{0pt}{1.2cm}
 $\displaystyle H^{\rm AE}=\frac{\mathbf{p}^2}{2m}+V$  & \makecell[c]{$\displaystyle\begin{aligned}
\mathcal{H}_{bj,b'i}^{\rm TB}=t^0_{bj,b'i}+\delta_{b'b}\delta_{ji}\Delta_{bi}    \end{aligned}$ }\\
\cline{1-2} \rule[-0.6cm]{0pt}{1.2cm}
  $\displaystyle \frac{\partial V}{\partial R_{s\alpha}}$  & \makecell[c]{$\displaystyle \frac{\partial \mathcal{H}^{\rm TB}_{bj,b'i}}{\partial R_{s\alpha}}$}\\
    \cline{1-2} \rule[-0.6cm]{0pt}{1.2cm}
$\displaystyle \frac{\partial^2 V}{\partial R_{s\alpha}\partial R_{s'\beta}}$ &   \makecell[c]{$\displaystyle   \frac{\partial^2 \mathcal{H}^{\rm TB}_{bj,b'i}}{\partial R_{s\alpha}\partial R_{s'\beta}} $} \\
\cline{1-2}
    \multicolumn{2}{c}{
     WITH NUCLEAR VELOCITY CORRECTIONS
    }\\
\cline{1-2}
    All-electron &  Tight-binding \\    
    \cline{1-2}    
       \rule[-0.5cm]{0pt}{1.5cm}  $\displaystyle H^{\rm AE}=\frac{\mathbf{p}^2}{2m}+V$ &
        \makecell[c]{
        $\displaystyle\begin{aligned}
           &\mathcal{H}_{bj,b'i}^{\rm TB}=\delta_{b'b}\delta_{ji}\Big(\Delta_{b'i}-\frac{m|\dot{\mathbf{R}}_{b'}(t)|^2}{2}\Big)+\frac{(1-\delta_{bb'})t_{bj,b'i}}{2}\Big( e^{\frac{i}{\hbar}m\dot{\mathbf{R}}_{b'}(t) \cdot (\mathbf{R}_{b}(t)-\mathbf{R}_{b'}(t))}+e^{\frac{i}{\hbar}m\dot{\mathbf{R}}_{b}(t) \cdot (\mathbf{R}_{b}(t)-\mathbf{R}_{b'}(t))}\Big)
        \end{aligned}$ \\
        (Eq. \eqref{eq:TB_hamiltonian})}\\

       \cline{1-2}\rule[-0.7cm]{0pt}{1.4cm}
       $\displaystyle \frac{\partial V}{\partial R_{s\alpha}}$ & 
       \makecell[c]{
       $\displaystyle \frac{\partial \mathcal{H}^{\rm TB}_{bj,b'i}}{\partial R_{s\alpha}}+\omega\frac{m}{2\hbar} \left(R_{b\alpha}- R_{b'\alpha}\right)\mathcal{H}^{\rm TB}_{bj,b'i}  (\delta_{bs}+\delta_{b's})\qquad$
         (Eq. \eqref{eq:eph_1st_TB_real_space})}
       \\
       \cline{1-2}\rule[-0.6cm]{0pt}{1.2cm}
       $\displaystyle \frac{\partial^2 V}{\partial R_{s\alpha}\partial R_{s'\beta}}$&
       \makecell[c]{
       $\displaystyle \begin{aligned}\\
    &\frac{\partial^2 \mathcal{H}^{\rm TB}_{bj,b'i}}{\partial R_{s\alpha}\partial R_{s'\beta}}+\omega\frac{m}{2\hbar} \left(R_{b\beta}- R_{b'\beta}\right)\frac{\partial \mathcal{H}^{\rm TB}_{bj,b'i}}{\partial R_{s\alpha}}(\delta_{bs'}+\delta_{b's'})-\omega\frac{m}{2\hbar} \left(R_{b
\alpha}- R_{b'\alpha}\right)\frac{\partial \mathcal{H}^{\rm TB}_{bj,b'i}}{\partial R_{s'\beta}}(\delta_{bs}+\delta_{b's})\\
&+\omega\frac{m}{\hbar}\mathcal{H}^{\rm TB}_{bj,b'i}\left( \delta_{bs}\delta_{b's'}-\delta_{bs'}\delta_{b's}\right) \delta_{\alpha\beta}\nonumber
\\&-\omega^2\delta_{ss'}\Bigg(m\delta_{bb'}\delta_{bs}\delta_{\alpha\beta}\delta_{ij}+\frac{m^2}{2\hbar^2}\left(R_{b\alpha}- R_{b'\alpha}\right)\left(R_{b\beta}- R_{b'\beta}\right)\mathcal{H}^{\rm TB}_{bj,b'i}(\delta_{bs}+\delta_{b's})\Bigg)
\end{aligned}$\\
(Eq.\eqref{eq:eph_2nd_TB_real_space})}\\
\cline{1-2}
    \end{tabular}
    \caption{Hamiltonians (first row) and first- and second-order nuclear displacement derivatives (second and third row respectively) with and without velocity-including corrections in the atomic orbitals for different levels of treatment of the electronic problem (all-electron and tight-binding). $m$ is the electronic mass. The potentials and the tight-binding hopping depend on the nuclear positions, although it is not explicitly indicated.}
    \label{tab:summary_eph}
\end{table}

    \begin{table}[h!]
\centering
    \begin{tabular}{|c||c|}
       \toprule
\multicolumn{1}{|c|}{NO CORRECTIONS} & \multicolumn{1}{c||}{ WITH NUCLEAR VELOCITY CORRECTIONS } \\ 
 \midrule
 \rule[-1.5cm]{0pt}{3cm}
\makecell[c]{
$H_\eta^{\mathrm{D}}=H_\eta^{\rm D,el}+H_\eta^{\rm D,e-ph}$\\
$
    \begin{aligned}
    H_\eta^{\rm D,el}=&\hbar v_F \Bigg(\eta p_x\sigma^\mathrm{P}_x+p_y\sigma^\mathrm{P}_y\Bigg)+\Delta(\eta)\sigma^\mathrm{P}_z,\\
    H_\eta^{\rm D,e-ph}=&\hbar v_F \xi_{\rm e-ph} \Bigg(-\left(u_{A,y}(t)-u_{B,y}(t)\right)\sigma^\mathrm{P}_x\\&
    +\eta \left(u_{A,x}(t)-u_{B,x}(t)\right)\sigma^\mathrm{P}_y\Bigg).
\end{aligned}$\\
(Eqs. \eqref{eq:low_energy} and \eqref{eq:low_energy_eph})
}
        &
\makecell[c]{
$H_\eta^{\mathrm{D}}=H_\eta^{\rm D,el}+H_\eta^{\rm D,e-ph}+
H_\eta^{\rm D,e-ph,\dot{R}}$\\
$ \begin{aligned}
    H_\eta^{\rm D,e-ph,\dot{R}}=&-\frac{1}{2}m v_F \Bigg( \eta \left(\dot{u}_{A,x}(t)+\dot{u}_{B,x}(t)\right)\sigma^\mathrm{P}_x\\&+\left(\dot{u}_{A,y}(t)+\dot{u}_{B,y}(t)\right)\sigma^\mathrm{P}_y\Bigg),
\end{aligned}
$\\
(Eq. \eqref{eq:low_energy_eph_dotR})}\\
   \multicolumn{2}{|c|}{\rule[-0.5cm]{0pt}{1cm}\makecell[c]{where $\mathbf{u}_{s\alpha}(t)=\mathbf{R}_{s\alpha}(t)-\mathbf{R}_{s\alpha}(t=0)$; $\Delta(\eta)=\frac{\Delta}{2}-\eta 3\sqrt{3}t_2$ for the Haldane model, $t_2=0$ for gapped graphene}}
   \\
\hline
 \rule[-0.5cm]{0pt}{1.5cm}
\makecell[c]{$\displaystyle \frac{\partial H^{\mathrm{D}}}{\partial u_{s\alpha}}$}    &
\makecell[c]{$\displaystyle \ \frac{\partial H^{\mathrm{D}}(\mathbf{p})}{\partial R_{s\alpha}}\to  \frac{\partial H^{\mathrm{D}}(\mathbf{p})}{\partial R_{s\alpha}}+\frac{1}{2}i\omega m v_{F}\sigma^\mathrm{P}_{\alpha}\left(\eta\delta_{\alpha x}+\delta_{\alpha y} \right)$ ,\qquad Eq. \eqref{eq:Dirac_1st_derivative_eph}} \\ 
     \bottomrule
    \end{tabular}
    \caption{In the first row: Dirac Hamiltonian describing the low-energy physics around the valley points ( $\rm \mathbf{K}$ and $\rm \mathbf{K}'$) of  crystals with honeycomb lattice and a diatomic basis, such as gapped graphene and for the Haldane model with the electron-phonon coupling term with and without nuclear velocity corrections. $\eta=\pm1$ in the valleys $\mathbf{K}$ and $\mathbf{K}'$ respectively. The electron-phonon coupling enters as a gauge field in the Hamiltonian. In the second row: first-order nuclear-displacement derivatives in the frequency space. The electrons in the Haldane model are spinless, whereas the bands of gapped graphene have a twofold spin degeneracy (neglecting the spin-orbit coupling).}
    \label{tab:Dirac_Hamiltonian_electron-phonon}
\end{table}

\end{widetext}

\clearpage

\section{Applications to gapped graphene and the Haldane model}
\label{sec:appl}
In this section, we investigate the nuclear velocity corrections to the tight-binding and low-energy vibrational responses of a 2D diatomic honeycomb crystal, tuning the parameters of the system in order to address the dependence of the correction on them and on the symmetries. 
We separate the nuclear displacement from the nuclear velocity contributions to the vibrational responses to clearly distinguish the standard expression (labelled with $\mathrm{R}$) from the corrections (labelled with $\mathrm{\dot{R}}$) 
\begin{align}
    &\mathcal{Z}^*_{\alpha,s\beta}(\omega)=\mathcal{Z}^{\mathrm{R}}_{\alpha,s\beta}(\omega)+\mathcal{Z}^{\mathrm{\dot{R}}}_{\alpha,s\beta}(\omega),\label{eq:correzioni_BEC}
    \end{align}
    \begin{align}
    \mathcal{C}_{s\alpha,s'\beta}(\omega)=\mathcal{C}^{\mathrm{R}}_{{s\alpha},{s'\beta}}(\omega)+\mathcal{C}^{\mathrm{\dot{R}}}_{{s\alpha},{s'\beta}}(\omega).\label{eq:correzioni_matrice_dinamic}
\end{align}
In tight-binding models, as discussed in Section \ref{sec:sum_rules}, the summation over the sublattices of $\displaystyle \mathcal{Z}^{\mathrm{R}}_{\alpha,s\beta}(\omega), \mathcal{C}^{\mathrm{R}}_{{s\alpha},{s'\beta}}(\omega)$  vanishes, while $\displaystyle \mathcal{Z}^{\mathrm{\dot{R}}}_{\alpha,s\beta}(\omega), \mathcal{C}^{\mathrm{\dot{R}}}_{{s\alpha},{s'\beta}}(\omega)$ yield non-zero contributions, corresponding to the related all-electron sum rules.

\paragraph*{Model}
We consider the Haldane model \cite{PhysRevLett.61.2015} - breaking both inversion and time-reversal symmetry - as the more general system 2D diatomic honeycomb crystal, obtaining graphene and gapped graphene as specific cases (adding a double spin degeneracy).  In the following sections, we use the low-energy model in the discussion since it provides more transparent physical insights into the results. The low-energy expression for the Haldane model around the valley $\eta$  ($\rm \mathbf{K}$ and $\rm \mathbf{K}'$ correspond to $\eta=\pm1$ respectively) with $\mathbf{p}=\mathbf{k}-\mathbf{K}(\mathbf{K}')$ is expressed in terms of  the Pauli matrices $\sigma^\mathrm{P}$, describing the sublattice degree of freedom of the two sites
\begin{equation}
    \begin{aligned}
    H_\eta^{\rm D,el}(\mathbf{p})&=\hbar v_F \Bigg(\eta p_x\sigma^\mathrm{P}_x+p_y\sigma^\mathrm{P}_y\Bigg)+\Delta(\eta)\sigma^\mathrm{P}_z,
    \label{eq:low_energy_Haldane}
\end{aligned}
\end{equation}
where $v_F$ is the Fermi velocity; $\Delta(\eta)=\frac{\Delta}{2}-\eta 3\sqrt{3}t_2$ for the Haldane model with $t_2$ imaginary second-nearest neighbour hopping, with the sign chosen according to the arrows shown in panel (a) of Figure \ref{fig:BEC_resonant_haldane}. Setting $t_2=0, \,\Delta(\eta)=\frac{\Delta}{2}$ and considering a double spin degeneracy,  the Hamiltonian describes gapped graphene, that compared to graphene ($\Delta=0$) has a different on-site energy on the two sublattices, mimicking a material like $\rm h-BN$ .  We adopt the parameters of graphene, i.e. $a=2.46$ \AA  (the unit cell area $V_c=\frac{\sqrt{3}}{2}a^2$) and $\hbar v_F=7.2 \mathrm{\,eV\cdot \text{\AA}}$ (corresponding to a first nearest neighbour hopping $t_1=3.4 \mathrm{\,eV}$) \cite{PhysRevB.84.035433}.  $\Delta$ and $t_2$ break the inversion and the time reversal symmetry respectively, enabling us to explore the dependence on these symmetries.  The system is a non-trivial Chern insulator for $|\Delta|<6\sqrt{3}|t_2|$ with $C=\pm1$. For $t_2, \Delta \neq0$ the $\mathbf{K}$ and $\mathbf{K}'$ points are inequivalent with two different gaps, denoted as $\Delta_1=|\Delta(\eta=1)|$ and $\Delta_2=|\Delta(\eta=-1)|$ in panel (b) in Figure \ref{fig:BEC_resonant_haldane}. The non-trivial topological state is characterised by band inversion at one of the two Dirac points, as plotted qualitatively in panel (b). 

As described in Section \ref{sec:kp_models} (see Eq. \eqref{eq:low_energy_eph} in particular), the electron-phonon coupling enters as a gauge field in the Hamiltonian, implying that nuclear displacement derivatives are related to the crystalline momentum derivatives of the Hamiltonian 
\begin{align}
\frac{\partial H_\eta^{\rm D,el}(\mathbf{p})}{\partial u_{sx}}=3l_s \eta  \frac{\beta_{\rm e-ph}}{a^2} {\frac{\partial H_\eta^{\rm D,el}(\mathbf{p})}{\partial p_y}},\\
\frac{\partial H_\eta^{\rm D,el}(\mathbf{p})}{\partial u_{sy}}= -3l_s \eta  \frac{\beta_{\rm e-ph}}{a^2} {\frac{\partial H_\eta^{\rm D,el}(\mathbf{p})}{\partial p_x}},
\end{align}
where $l_s=\pm1$ for the sites with $\pm \frac{\Delta}{2}$, respectively. In the following, we adopt the dimensionless electron-phonon coupling parameter  $\displaystyle \beta_{\rm e-ph}=2.38$ obtained from the \textit{ab initio} electron-phonon matrix element evaluated between the valence and conduction band and averaged over the degenerate optical phonon at the zone centre of graphene. $\displaystyle \beta_{\rm e-ph}$ corresponds to $\displaystyle \xi_{\mathrm{e-ph}}=3\frac{\beta_{\rm e-ph}}{a^2}$, introduced in Eq. \eqref{eq:low_energy_eph}. With the Dirac low-energy Hamiltonian, the nuclear velocity correction to the electron-phonon coupling (Eq. \eqref{eq:Dirac_1st_derivative_eph} and Table \ref{tab:Dirac_Hamiltonian_electron-phonon}) yields a simple expression in terms of the band velocity.  

In the following Sections we discuss the nuclear velocity correction to the non-adiabatic vibrational responses in the tight-binding and low energy model for a 2D diatomic honeycomb crystal, tuning the systems' parameters. In detail, Subsections \ref{subsec:BEC_metals} and \ref{subsec:BEC_Haldane} focus on the corrections to the Born effective charges; Subsection \ref{subsec:phonon_lifetime} on the phonon properties. Specifically, in Subsection \ref{subsec:BEC_metals}, the correction to the Born effective charges in doped gapped graphene is benchmarked against \textit{ab initio} calculations, also accounting for the nuclear velocity correction in the pseudopotentials \cite{PhysRevLett.136.196401}. The case when phonons resonate with electronic interband transitions is discussed in Subsection \ref{subsec:BEC_Haldane}, both for the (insulating) gapped graphene and for the Haldane model, addressing the role of time-reversal symmetry breaking and of non-trivial electronic topological states. Finally, in Subsection \ref{subsec:phonon_lifetime}, for gapped graphene we study the correction of the phonon frequency and linewidth, varying the atoms' masses and addressing the electron-phonon coupling dependence of the correction.

\subsection{Born effective charges in metallic systems and at resonance between electronic and vibrational excitations} \label{subsec:BEC_metals}
\begin{figure*}[h]
    \centering
    \includegraphics[width=0.9\textwidth]{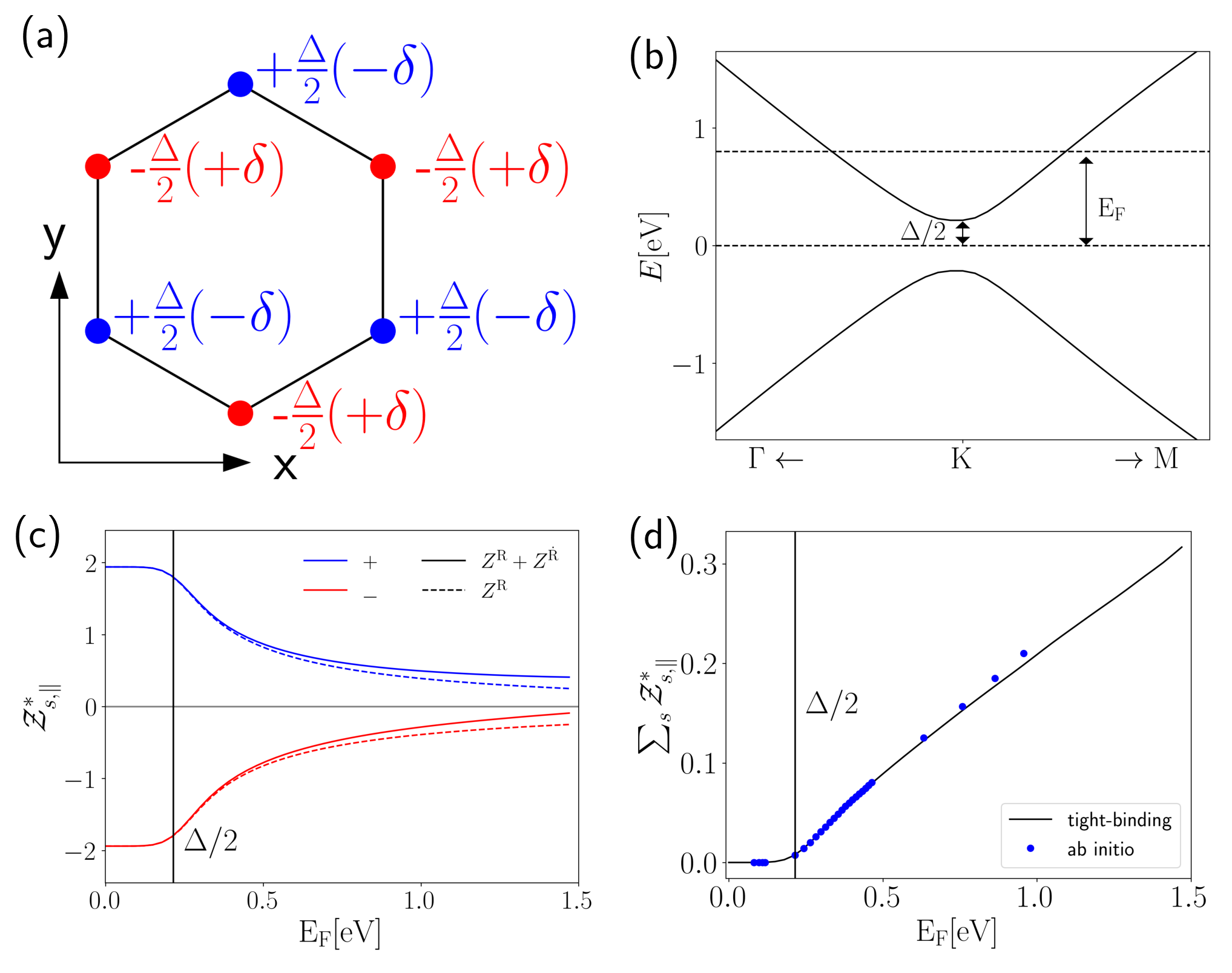}
    \caption{(a) Gapped graphene lattice model, with alternating sites with either an excess or a deficiency of electric charge, corresponding to negative and positive on-site energies in the tight-binding model, respectively. (b) Energy bands of gapped graphene near the $\mathbf{K}$ point along the high symmetry reciprocal space path $\rm \Gamma \to K \to M$. The Fermi level is set at zero in the middle of the gap. (c) In-plane component of the Born effective charges at $\omega=i0^+$ of gapped graphene, as a function of the Fermi level computed via the tight-binding model corrected with non-inertial effects. Blue lines are for the + atom, red ones for the - one. Continuous lines indicate Born effective charges corrected by velocity terms ($\mathcal{Z}^{\rm R}+\mathcal{Z}^{\rm \dot{R}}$), while dashes line indicate uncorrected ones ($\mathcal{Z}^{\rm R}$). The vertical black dashed line indicates the Fermi level corresponding to the bottom of the conduction band $E_F=\Delta/2$. (d) Comparison of the sum of the diagonal components of the metallic Born effective charges computed with the tight-binding model (black line) and with \textit{ab initio} simulations (blue dots), as a function of the Fermi level. Both the \textit{ab initio} simulations and tight-binding calculations are performed with a Gaussian smearing corresponding to a temperature of $T=27 \mathrm{ meV}$.}
    \label{fig:BEC_sum_rule}
\end{figure*}
The frequency-dependent Born effective charges tensor of 2D honeycomb diatomic crystal has equal diagonal in-plane components and opposite off-diagonal components $\displaystyle \mathcal{Z}^*_{s,xy}(\omega)=-\mathcal{Z}^*_{s,yx}(\omega)$ because of the $C_3$ symmetry \cite{Bistoni2019, PhysRevB.110.L201405, Fachin2025}. Time-reversal symmetry constrains the off-diagonal components to be zero due to the Onsager relations, whereas in its absence those terms are allowed - as it happens also for the off-diagonal components of the optical conductivity  \cite{Resta2022, 10.21468/SciPostPhysCore.5.3.039}. Therefore, for the time-reversal invariant gapped graphene, we present the only independent in-plane component per atom  $\displaystyle \mathcal{Z}^*_{s,\parallel}$, while for the time-reversal breaking Haldane model also the off-diagonal component.  The out of plane $\mathcal{Z}^*_{s,zz}$ component of the Born effective charges corresponds only to the rigid translation of the electric charge on the site, without any contribution due to the phonon perturbation on the electronic polarisation. This is a consequence of the mirror reflection symmetry with respect to the plane of the 2D material that forbids low-energy excitations originating from perpendicular perturbation in the linear response regime \cite{PhysRevB.103.134304, Bistoni2019}. In a tight-binding approach, the $\displaystyle \mathcal{Z}^*_{s,zz}$ coincide with the Mulliken charge on the atomic site \cite{Mulliken, PhysRevB.58.6224, PhysRevB.110.L201405}.

In the Dirac low-energy model, the nuclear velocity contribution to the Born effective charges is equal on the two sublattices, coinciding with half of the sum rule on each
\begin{align}
    \mathcal{Z}^{\mathrm{\dot{R}}}_{\alpha,s\beta}(\omega)=-i\omega \frac{ m}{2e^2}V_c\sigma_{\alpha\beta}(\omega).
\end{align}
The correction is enhanced by a weak electron-phonon coupling since $\displaystyle \mathcal{Z}^{\mathrm{R}}_{\alpha,s\alpha}(\omega)\propto \beta_{\mathrm{e-ph}}$ while $\mathcal{Z}^{\mathrm{\dot{R}}}_{\alpha,s\alpha}(\omega)$ does not depend on it.

In the following, we consider two cases where the non-adiabatic effects are relevant, a metal and an insulator with a gap resonating with the phonon frequency. 

\subsubsection{Metallic gapped graphene}
For metallic systems, even in the zero-frequency limit, the sum rule for the frequency-dependent Born effective charges does not vanish \cite{Marchese2024, PhysRevLett.128.095901, PhysRevB.107.094308, PhysRevB.110.094306}. This originates from the presence of conducting electrons, which contribute to the DC conductivity. In the infinite electronic lifetime (clean) limit, the intraband (Drude) contribution to the optical conductivity is  
\begin{equation}
    \begin{aligned}
        (\sigma_{\alpha\beta})^{\mathrm{D}}(\omega)=& \frac{(\omega_p^*)^2_{\alpha\beta}}{4} \delta( \omega)+i\frac{(\omega_p^*)^2_{\alpha\beta}}{4\pi}\frac{1}{ \omega}
    \end{aligned}
    \label{eq:sigma_drude}
\end{equation}
where $(\omega_p^*)^2_{\alpha\beta}$ is the effective plasma frequency. In the limit for $\omega \to 0$ the imaginary part of the intraband optical conductivity plugged into the Born effective charges sum rule (Eq. \eqref{eq:BEC_sum_rule}) gives \cite{Marchese2024, PhysRevLett.128.095901, Resta2022}
\begin{equation}
   \sum_s \mathcal{Z}^*_{\alpha,s\beta}(\omega=i0^+)=\frac{m V_{\rm c}}{e^2}\frac{(\omega_p^*)^2_{\alpha\beta}}{4\pi}.
   \label{eq:BEC_metallic_sum_rule}
\end{equation}
Gapped graphene is described with a tight-binding model as in Ref.\cite{Bistoni2019}. The tight-binding electron-phonon coupling is introduced as the variation of hopping due to a change in the bond length, as described in detail in Refs. \cite{Fachin2025, fachin2026lattice, PhysRevB.84.035433, Bistoni2019}. The corresponding nuclear velocity corrections are summarised in Table \ref{tab:summary_eph}.  In the tight-binding framework, doping is introduced in gapped graphene by rigidly shifting the Fermi level from the middle of the gap (see panel (b) of Fig. \ref{fig:BEC_sum_rule}).  We compare tight-binding results for the Born effective charges to the \textit{ab initio} calculations, performed using a modified version of Quantum Espresso \cite{Giannozzi2017}, already used and discussed in Ref. \cite{Marchese2024}. Gapped graphene is obtained by generating pseudopotentials with the valence charges of the two atoms $Z_{+}=6-\delta$ and $Z_{-}=6+\delta$ using the \emph{atomic} package --- as in Ref. \cite{Bistoni2019} --- while doping is introduced with a gate potential, as described in Ref. \cite{PhysRevB.89.245406}. A tight-binding positive on-site energy $+\frac{\Delta}{2}$ corresponds to a negative variation of the valence charge $-\delta$, as indicated in panel (a) of Fig. \ref{fig:BEC_sum_rule}.  In our calculations, we set a difference in the valence charge of the two atoms of $2\delta=0.2$ that produces a direct gap at the $\mathbf{K}$ point $E_{\mathrm{gap}}=0.43 \mathrm{\, eV}$- corresponding to $\Delta$ in the tight-binding approach - as plotted in panel (b) of Fig. \ref{fig:BEC_sum_rule}. In the \textit{ab initio} simulation, the Drude weight is computed using two velocity vertices to include the nuclear velocity correction to the pseudopotentials \cite{PhysRevLett.136.196401}. Nevertheless, for the studied system, the correction is negligible, unlike other systems, such as $\rm H_3S$ or $\rm Al$, where a $5-10\%$ difference is observed \cite{Marchese2024, Marchese2024_phdthesis}. Both computations are performed with a Gaussian smearing of $\sim 0.068 \rm \,eV$, corresponding to room temperature in the Fermi statistics \cite{PhysRevB.107.195122}. 

According to the symmetry analysis in doped gapped graphene, the Born effective charges tensor of gapped graphene is diagonal with equal real-valued in-plane components $\displaystyle \mathcal{Z}^*_{s,\parallel}$, plotted in panels (c-d) of Fig. \ref{fig:BEC_sum_rule}. As the Fermi level is tuned from the reference value placed in the middle of the gap (see panel (b) of Fig. \ref{fig:BEC_sum_rule}) - the Born effective charges decrease while the doping level becomes larger, as shown in panel (c) of Fig. \ref{fig:BEC_sum_rule}. While $\mathcal{Z}^{\mathrm{R}}$ is equal and opposite on the two sites, $\mathcal{Z}^{\mathrm{\dot{R}}}$ shifts both the Born effective charges on the two sites by the same amount, making them asymmetric about zero.  The correction depends significantly on the doping level, reaching values up to $50\%$ for large doping levels ($E_F \approx 1.5 \mathrm{ \,eV}$). $\mathcal{Z}^{\mathrm{\dot{R}}}$  makes the sum over the two sites of the Born effective charges no longer vanish as in the undoped semiconductor case, as plotted in panel (d) of Figure \ref{fig:BEC_sum_rule}. The \textit{ab initio} Drude weight - representing the sum rule according to Eq. \eqref{eq:BEC_metallic_sum_rule} - is in excellent agreement with the tight-binding sum rule over the in-plane component of the Born effective charges.  Summarising, the inclusion of the nuclear velocity-dependent phase originating from ionic motion in the tight-binding approach enables the recovery of the all-electron behaviour both from a qualitative and quantitative point of view, as revealed by the excellent agreement in the sum rule with \textit{ab initio} simulation.

\subsubsection{Resonance between electronic and vibrational excitations in gapped graphene and Haldane model}
\label{subsec:BEC_Haldane}
In this section, we study the nuclear velocity correction to the Born effective charges when the phonon frequency resonates with electronic interband transitions for the time-reversal invariant (gapped graphene) and breaking (Haldane model) systems, within the low-energy Dirac model. The explicit expressions for the Born effective charges and the optical conductivity for gapped graphene are reported in Ref. \cite{Bistoni2019, Fachin2025} and for the Haldane model in Ref. \cite{fachin2026lattice}. In the following, we set as the phonon frequency the graphene degenerate optical phonon at the zone centre, $\omega_{\mathrm{ph}}=0.2 \mathrm{\,eV}$.

\paragraph{Gapped graphene}
\begin{figure*}[h!]
    \centering
    \includegraphics[width=0.99\linewidth]{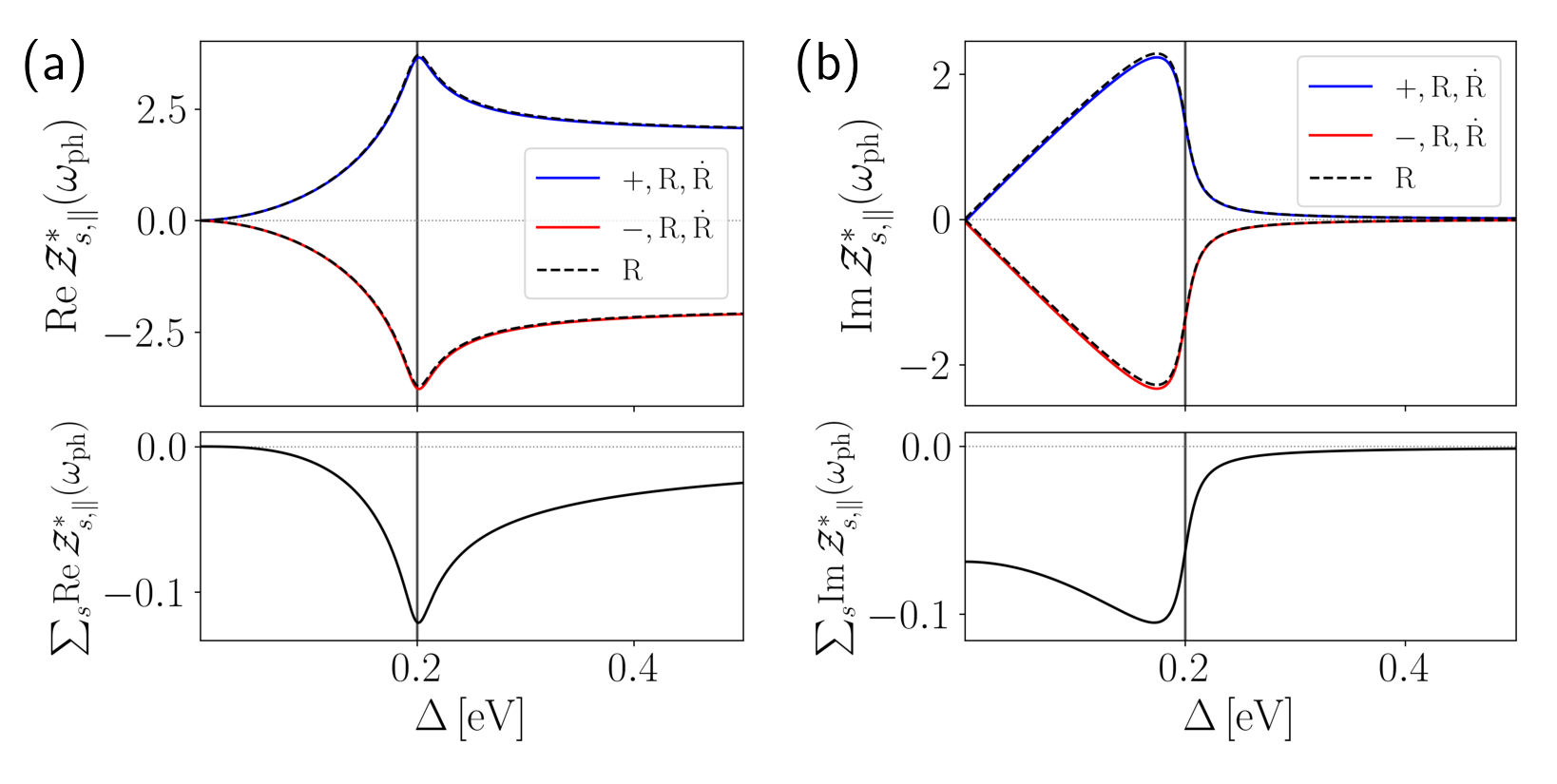}
    \caption{ Real (a) and imaginary (b) parts of the Born effective charges and the related sum rules are plotted as a function of the electronic band gap $\Delta$ for gapped graphene. The black dotted lines are the Born effective charges without the nuclear velocity corrections. The vertical gray lines correspond to the phonon frequency $\omega_{\mathrm{ph}}=0.2 \mathrm{\,eV}$. }
    \label{fig:BEC_resonant}
\end{figure*}
According to the symmetry analysis, the Born effective charges tensor for gapped graphene is diagonal with equal in-plane components. From Eq. \eqref{eq:correzioni_BEC}, within the low-energy model for graphene, the correction is 
\begin{equation}
    \mathcal{Z}^{\mathrm{\dot{R}}}_{\alpha,s\alpha}(\omega)=-i\hbar\omega \frac{ m}{8\hbar^2}V_c\frac{\sigma_{\alpha\alpha}(\omega)}{\sigma_0}
    \label{eq:correzioni_BEC_gapped_graphene}
\end{equation}
where $\sigma_0=\frac{e^2}{4\hbar}$ and $\displaystyle\frac{ m}{8\hbar^2}V_c=0.085 \mathrm{\,eV}^{-1}$. Since the system is an insulator, the optical conductivity has only interband contributions, computed as in Refs. \cite{Bistoni2019, Fachin2025}. In Figure \ref{fig:BEC_resonant},  the real and imaginary part of $\mathcal{Z}^*_{s, \parallel}(\omega_{\mathrm{ph}})$ are plotted as a function of the band gap, continuously tuned from $\Delta << \omega_{\mathrm{ph}}$ to the electronically off-resonant condition ( $\Delta >>\omega_{\mathrm{ph}}$). In the real part of $\mathcal{Z}^*_{s, \parallel}$ (panel (a)) the correction - proportional to $\displaystyle \omega_{\mathrm{ph}}\mathrm{Im}\left(\sigma_{\alpha\alpha}(\omega_{\mathrm{ph}})\right)$ and corresponding to half the sum (bottom panel of (a)) for each site - peaks at the resonance where it is approximately $\sim 2\%$. A similar behaviour is observed also for the imaginary part, where the correction is proportional to $\displaystyle \omega_{\mathrm{ph}}\mathrm{Re}\left(\sigma_{\alpha\alpha}(\omega_{\mathrm{ph}})\right)$.
Larger phonon frequencies, small electron-phonon coupling and large optical conductivity enhance the correction.

\begin{figure*}[h!]
    \centering
    \includegraphics[width=0.99\linewidth]{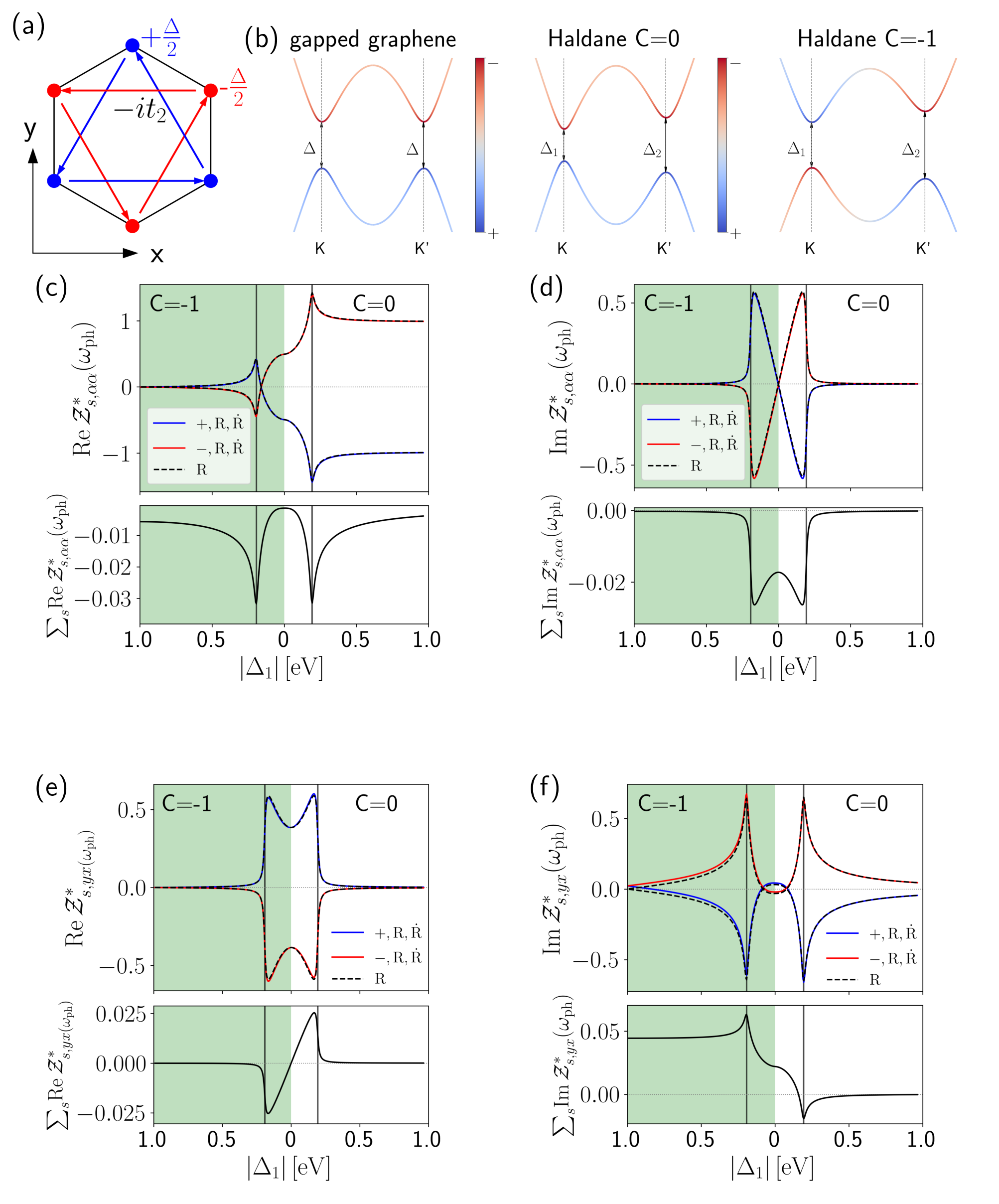}
    \caption{(a)Lattice representation of the Haldane model. Arrows identify the direction for the $\rm 2^{nd}$ nearest neighbours hopping. (b) Qualitative sketch of energy bands along the high symmetry direction for gapped graphene, and for the Haldane model in the trivial case (C=0) and non trivial states (C=1). 
   (c-f) 
   Real and imaginary parts of the $\mathcal{Z}^*_{s, \alpha\alpha}(\omega_{\mathrm{ph}})$ and $\mathcal{Z}^*_{s, yx}(\omega_{\mathrm{ph}})$ as a function of the gap at the $\mathbf{K}$ point $\Delta_1$ by varying the on-site energy $\Delta$ and keeping $t_2=0.1 \mathrm{\,eV}$ and tuning the system from a topological (green region) to a trivial state. The dotted black lines are the Born effective charges without the nuclear velocity-correction. The vertical gray lines correspond to the phonon frequency of $\omega_{\mathrm{ph}}=0.2 \mathrm{\,eV}$.}
    \label{fig:BEC_resonant_haldane}
\end{figure*}

\paragraph{Haldane model}
As discussed in the symmetry analysis, the time-reversal symmetry breaking enables also non-zero off-diagonal components of the Born effective charges  tensor. The diagonal terms $\displaystyle \mathcal{Z}^*_{s, \alpha\alpha}(\omega_{\mathrm{ph}})$ are plotted in panels (c-d) of Figure \ref{fig:BEC_resonant_haldane} as a function of the electronic band gap $\Delta_1$ across the non-trivial (corresponding to green regions) and trivial states, tuned by varying the on-site energy $\Delta$ and keeping $t_2=0.1 \mathrm{\,eV}$. The electronically off-resonant limit of $\mathcal{Z}^*_{s, \alpha\alpha}$ in the non trivial and trivial states are completely different. The behaviour in the trivial state corresponds to half the value of gapped graphene (due to the double spin degeneracy), according to Figure \ref{fig:BEC_resonant}.  Indeed, in diatomic 2D honeycomb crystals, the dominant contributions to the Born effective charges originate from the valleys $\mathbf{K}$ and $\mathbf{K}'$.  In trivial insulators (with and without time-reversal symmetry) these contributions from the two valleys sum up in the diagonal terms $\displaystyle \mathcal{Z}^{\mathrm{R}}_{s, \alpha\alpha}$\cite{Bistoni2019}. Conversely, in a topological non trivial state, owing to the topological band inversion (see panel (b) of Fig. \ref{fig:BEC_resonant_haldane}), they are opposite, yielding a nearly vanishing zero-frequency limit \cite{PhysRevB.110.L201405}.   The nuclear velocity correction to $\displaystyle \mathcal{Z}^*_{s, \alpha\alpha}(\omega)$ is the same as gapped graphene, given in Eq. \eqref{eq:correzioni_BEC_gapped_graphene}, which does not depend on the electronic topological state, as shown in the bottom Figures of panels (c-d) of Fig. \ref{fig:BEC_resonant_haldane}. The correction is peaked around the resonance, remaining small as in gapped graphene. In the non-trivial state, the relative correction is larger, due to the smaller value of $\displaystyle \mathcal{Z}^{\mathrm{R}}_{s, \alpha\alpha}(\omega_{\mathrm{ph}})$. 

For the off-diagonal component, the nuclear velocity correction reads
\begin{equation}
    \mathcal{Z}^{\mathrm{\dot{R}}}_{x,sy}(\omega)=-i\hbar\omega \frac{ m}{8\hbar^2}V_c\frac{\sigma_{xy}(\omega)}{\sigma_0},
    \label{eq:correzioni_BEC_Haldane}
\end{equation}
where $\displaystyle \sigma_{xy}(\omega)$ is the frequency-dependent antisymmetric Hall conductivity.  While $\displaystyle \mathcal{Z}^{\mathrm{R}}_{s, xy}(\omega)$ does not depend on the topological state, $ \mathcal{Z}^{\mathrm{\dot{R}}}_{x,sy}(\omega)$ does, being proportional to the Hall conductivity.  Such behaviour of $\displaystyle \mathcal{Z}^{\mathrm{R}}_{s, xy}$ is explained by the electron-phonon entering as a gauge field at the valleys, a detailed discussion is presented in Ref. \cite{fachin2026lattice}. The nuclear velocity correction to $\displaystyle \mathcal{Z}^{*}_{s, xy}(\omega_{\mathrm{ph}})$ are shown in panels (e-f) of Fig. \ref{fig:BEC_resonant_haldane}. In particular,  the correction to $ \mathrm{Im}\left(\mathcal{Z}^*_{x,sy}(\omega_{\mathrm{ph}})\right)$ - depending on $\omega_{\mathrm{ph}}\mathrm{Re}\left(\sigma_{xy}(\omega_{\mathrm{ph}})\right)$ - remains finite in the non-trivial state even in the electronically off-resonant limit where it is $\omega_{\mathrm{ph}}\mathrm{Re}\left(\sigma_{xy}(\omega_{\mathrm{ph}})\right)=\omega_{\mathrm{ph}}\sigma_0^{\mathrm{AHC}}$, with $\sigma_0^{\mathrm{AHC}}=\frac{e^2}{h}$ . 

\subsection{Phonon frequency and linewidth of gapped graphene}
\label{subsec:phonon_lifetime}
In this Section, we discuss the correction to the phonon frequency and lifetime for gapped graphene at the zone centre, varying the atoms' masses and addressing the dependence on the electron-phonon coupling, within the low-energy model. The nuclear velocity correction is expressed for this model as
\begin{align}
   \mathcal{C}^{\mathrm{\dot{R}}}_{{s\alpha},{s'\beta}}(\omega)=\frac{m\omega^2}{2}\Big[\mathcal{Z}^{\mathrm{el}}_{\beta,s\alpha}(\omega)+
   \mathcal{Z}^{\mathrm{el}}_{\alpha,s'\beta}(\omega)-2\delta_{ss'}\delta_{\alpha\beta} \rho_s\Big].
\end{align}
where $\mathcal{Z}^{\mathrm{el}}$ is the electronic part of the Born effective charges, as defined in Eq. \eqref{eq:BEC_definition_ionic_electronic}, $m$ the electron mass, $\rho_s$ the electronic charge density on the $s$ site.

The force constant matrix $\displaystyle\mathcal{C}^{\mathrm{R}}_{{s\alpha},{s'\beta}}(\omega)$ is obtained by adding the tight-binding self-energy contribution $\displaystyle \Pi^{\mathrm{R}}_{{s\alpha},{s'\beta}}(\omega)$ to the adiabatic force constant matrix $\displaystyle \mathcal{C}_{s\alpha,s'\beta}(\omega=0)$, built using a spring model as in Ref. \cite{PhysRevB.105.064303, fachin2026lattice} with parameters fitted to match the \textit{ab initio} frequencies computed with PHonon package of Quantum Espresso for  graphene. We focus on the in-plane doubly degenerate optical phonon at the zone centre, $\omega_{\rm ph}=0.2 \mathrm{ \,eV}$.

In a diatomic 2D honeycomb lattice, the $C_3$ and the time-reversal symmetry impose that the frequency-dependent force constant matrix is diagonal in cartesian indices - i.e. only $\displaystyle\mathcal{C}_{{s\alpha},{s'\alpha}}(\omega)$ are different from zero - with $\displaystyle\mathcal{C}_{{sx},{s'x}}(\omega)=\mathcal{C}_{{sy},{s'y}}(\omega)$. In addition, at $\omega=0$ the acoustic sum rule for a diatomic lattice imposes that  $\displaystyle \mathcal{C}_{{s\alpha},{s\alpha}}(\omega=0)=- \mathcal{C}_{{s\alpha},{s'\alpha}}(\omega=0)$.  As a consequence,  in gapped graphene, with atomic masses $M_C$, at the zone centre, the only independent component of the adiabatic force constant matrix is  $\displaystyle \mathcal{C}_{{s\alpha},{s\alpha}}(\omega=0)=M_C\frac{\omega_{\rm ph}^2}{2}$. Although the phonon self-energy $\displaystyle \Pi^{\mathrm{R}}_{s\alpha, s'\alpha}$ could have three different independent values by symmetry ($ \displaystyle \Pi^{\mathrm{R}}_{A\alpha, A\alpha},\Pi^{\mathrm{R}}_{A\alpha, B\alpha},\Pi^{\mathrm{R}}_{B\alpha, B\alpha} $ ),  because of the electron-phonon coupling of the low-energy model, these three elements have the same absolute value and they are equal to \cite{Bistoni2019, Fachin2025, fachin2026lattice}
\begin{equation}
    \Pi^{\mathrm{R}}_{s\alpha, s'\alpha}(\omega)=l_sl_{s'} \frac{9}{4}V_{\rm c}\left(\frac{\beta_{\mathrm{e-ph}}}{a^2}\right)^2\hbar\omega \frac{i\sigma_{\alpha\alpha}}{\sigma_0},
    \label{eq:self-energy_gapped_graphene}
\end{equation}
where $l_s=\pm1$ depending on the site. 

\paragraph*{Correction to the phonon frequency}
The phonon frequencies are determined by the eigenvalues of the Hermitian part of the dynamical matrix, according to Eq. \eqref{eq:ph_freq}, that coincide with the real part at the zone centre. Therefore, here we focus on the real part of the force constant matrix, where the dominant contribution derives from the adiabatic part $\displaystyle \mathcal{C}_{{s\alpha},{s'\alpha}}(\omega=0)$. The relative correction to the real part of the force constant matrix is 
\begin{equation}
\frac{ \mathcal{C}^{\mathrm{\dot{R}}}_{{s\alpha},{s'\alpha}}(\omega_{\mathrm
ph})}{|\mathcal{C}_{{s\alpha},{s'\alpha}}(\omega=0)|}=\frac{m\Big[\mathcal{Z}^{\mathrm{el}}_{\alpha,s\alpha}(\omega)+
   \mathcal{Z}^{\mathrm{el}}_{\alpha,s'\alpha}(\omega)-2\delta_{ss'}\rho_s\Big]}{M_C}
\end{equation}
where $\displaystyle \frac{m}{M_C}=4.6 \times10^{-5}$. Since the Born effective charges are usually of the order of unity, the correction is small, dominated by the renormalization of the nuclear mass with the electronic ones, especially for atoms with a large number of electrons, as pointed out clearly in the \textit{ab initio} framework by Ref. \cite{PhysRevLett.136.196401}.   
\paragraph*{Correction to the phonon lifetime}
By exploiting the symmetries of $\Pi^{\mathrm{R}}_{s\alpha, s\alpha}$ in a diatomic 2D honeycomb crystal with time reversal symmetry, Eq. \eqref{eq:ph_lifetime}, evaluated for the optical phonon modes at the zone centre, gives, for the electron-phonon coupling contribution to the phonon lifetime
\begin{equation}
    \gamma^{\mathrm{R}}_{\mathrm{e-ph}}=\frac{2\hbar^2}{M_C\hbar\omega_{\mathrm{ph}}} \mathrm{Im}\left( \Pi^{\mathrm{R}}_{s\alpha, s\alpha}(\omega_{\mathrm
ph})\right),
\end{equation}
that, by using Eq. \eqref{eq:self-energy_gapped_graphene}, explicitly is 
\begin{equation}
    \gamma^{\mathrm{R}}_{\mathrm{e-ph}} = \bar{\gamma}\beta_{\rm e-ph}^2  \mathrm{Re}\left(\frac{\sigma _{\alpha\alpha}(\omega_{\mathrm
ph})}{\sigma_0}\right).
\end{equation}
with $\displaystyle \bar{\gamma}=\hbar^2\frac{9\sqrt{3}}{4Ma^2}=0.22 \mathrm{\,meV}$ \cite{Bistoni2019, Fachin2025, fachin2026lattice}. 

\begin{figure*}
    \centering
    \includegraphics[width=\linewidth]{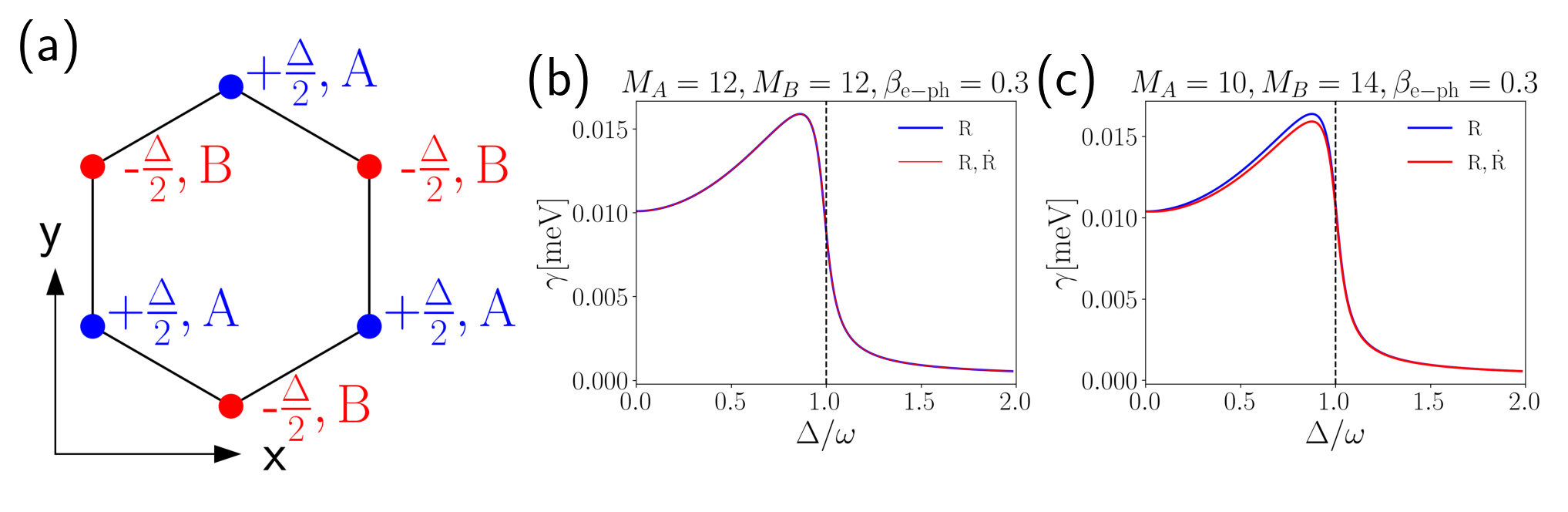}
    \caption{Phonon linewidth as a function of the ratio between the electronic band gap and the phonon frequency for gapped graphene (the lattice is represented in the panel (a)) with $\beta_{\mathrm{e-ph}}=0.3$. The case of equal masses - appropriate for graphene with $M_A=M_B=12 \mathrm{\, amu}$- is plotted in panel (b), whereas those of different masses - capturing the mass asymmetry of $\rm h-BN$ with  $M_A=14 \mathrm{\, amu}$ and $M_B=10 \mathrm{\, amu}$ - in the panel (c).}
    \label{fig:lifetime}
\end{figure*}

Since $\displaystyle \mathrm{Im}C_{s\alpha, s'\beta}= \mathrm{Im}\Pi_{s\alpha, s'\beta}$, here we consider the nuclear velocity correction in comparison to $\displaystyle \mathrm{Im}\Pi^{\mathrm{R}}_{s\alpha, s\alpha}$, where the latter is the only independent component of $\displaystyle \mathrm{Im}\Pi^{\mathrm{R}}_{s\alpha, s'\beta}$according to the symmetries of the model. The correction is different for $s=s'$ and $s\neq s'$. In the first case, the relative correction to $\displaystyle \mathrm{Im}\left(\mathcal{C}_{{s\alpha},{s\alpha}}(\omega)\right)$ is
\begin{equation}
    \frac{\mathrm{Im}\left(\mathcal{C}^{\mathrm{\dot{R}}}_{{s\alpha},{s\alpha}}(\omega)\right)}{\mathrm{Im}\left(\mathcal{C}^{\mathrm{R}}_{{s\alpha},{s\alpha}}(\omega)\right)}= \frac{4}{9}\frac{m}{\hbar^2}\frac{\hbar\omega}{V_{\rm c}\left(\frac{\beta_{\rm e-ph}}{a^2}\right)^2}\frac{\mathrm{Im} \left(\mathcal{Z}^{\mathrm{el}}_{{\alpha},{s\alpha}}(\omega)\right)}{\frac{\mathrm{Re}\left(\sigma_{\alpha\alpha}(\omega)\right)}{\sigma_0}},
    \label{eq:correction_imaginary_part_diagonal}
\end{equation}
where $\displaystyle\frac{m}{\hbar^2 }=0.13\, \mathrm{eV}^{-1}\text{\AA}^{-2}$. Since $\mathcal{Z}^{\mathrm{R}}$ scale as $\beta_{\mathrm{e-ph}}$, the correction scales as $\omega_{\mathrm{ph}}/\beta_{\mathrm{e-ph}}$, becoming larger for small electron-phonon coupling. 
In the second case ($ s\neq s'$), the contributions to the correction from $\mathcal{Z}^{\mathrm{R}}$ cancel, leaving only those from $\mathcal{Z}^{\mathrm{\dot{R}}}$. It follows that, since $\displaystyle\mathrm{Im}\left(\mathcal{Z}^{\mathrm{\dot{R}}} \right) \propto \mathrm{Re}\left(\sigma\right)$, 
\begin{equation}
\frac{\mathrm{Im}\left(\mathcal{C}^{\mathrm{\dot{R}}}_{{s\alpha},{s'\alpha}}(\omega)\right)}{\mathrm{Im}\left(\mathcal{C}^{\mathrm{R}}_{{s\alpha},{s'\alpha}}(\omega)\right)}=\frac{1}{18}\left( \frac{m}{\hbar^2}\right)^2\frac{(\hbar\omega)^2}{\left(\frac{\beta_{\rm e-ph}}{a^2}\right)^2}.
\end{equation}
By substituting the specific values of graphene for the optical phonon frequency, the electron-phonon coupling and the lattice constant, 
\begin{align}
\frac{\mathrm{Im}\left(\mathcal{C}^{\mathrm{\dot{R}}}_{{s\alpha},{s\alpha}}(\omega)\right)}{\mathrm{Im}\left(\mathcal{C}^{\mathrm{R}}_{{s\alpha},{s\alpha}}(\omega)\right)}=1.4\cdot 10^{-2}\frac{\mathrm{Im} \left(\mathcal{Z}^{\mathrm{el}}_{{\alpha},{s\alpha}}(\omega)\right)}{\frac{\mathrm{Re}\left(\sigma_{\alpha\alpha}(\omega)\right)}{\sigma_0}},\\
\frac{\mathrm{Im}\left(\mathcal{C}^{\mathrm{\dot{R}}}_{{s\alpha},{s'\alpha}}(\omega)\right)}{\mathrm{Im}\left(\mathcal{C}_{{s\alpha},{s'\alpha}}(\omega)\right)}=2.4\cdot 10^{-4}.
\end{align}
The nuclear velocity corrections to the phonon lifetime are shown in Figure \ref{fig:lifetime} for gapped graphene but with  $\beta_{\rm e-ph}=0.3$, smaller than the value of graphene ($\beta^{\rm graphene}_{\rm e-ph}=2.38$). Panel (b) of Figure \ref{fig:lifetime} shows that, for equal masses of the two sites, the nuclear velocity correction to the phonon lifetime remains negligible, also when the band gap $\Delta$ is tuned to resonate with the phonon frequency. Conversely, inequivalent masses enhance the effects on the phonon lifetime at the resonance. This is shown in the panels (c) of Figure \ref{fig:lifetime}, where the masses are chosen as $M_A=14\mathrm{\, amu}$ and $M_B=10\mathrm{\, amu}$, capturing the mass asymmetry of systems like h-BN (keeping all the other parameters fixed). Here, at the resonance between the lattice and the electronic interband excitations, for $\beta_{\rm e-ph}=0.3$, a $\sim 2-3\%$ correction is observed.

Summarising, the nuclear velocity corrections restores the all-electron sum rule for the force constant matrix. The corresponding correction to the phonon lifetime are enhanced for an imbalance between the nuclear masses involved in the phonon displacement. Moreover, the relative correction scales as the inverse of the electron-phonon coupling, becoming larger in the case of small electron-phonon coupling. In Dirac-like model, the correction is expressed in terms of the Born-effective charges, implying that it can be relevant in systems with giant Born effective charges, as, for instance, in the 1D Rice-Mele model discussed in Ref. \cite{Villani2024}.

\section{Conclusion}
\label{sec:concl}
We have shown a procedure to include the effect of the nuclear motion on atomic orbitals in effective models built using localised atomic orbitals -  such as LCAO and tight-binding models - and in low-energy models, focusing on the Dirac Hamiltonians for graphene and Haldane-like systems. In these models, the inclusion of nuclear-velocity dependent phases in the atomic orbitals is needed for describing non-adiabatic vibrational responses. In the LCAO approach, these phases affect the Ehrenfest dynamics through nuclear velocity and acceleration derivatives.  In tight-binding Hamiltonians, Peierls-like nuclear-velocity-dependent phases appear in the non-local potentials, therefore modifying both the nuclear Ehrenfest dynamics and the vibrational responses of the system. The correction to the electron-phonon coupling has a physically intuitive expression in low-energy Dirac Hamiltonians, where it is proportional to the phonon frequency and to the band velocity. In other words, a velocity vertex is added to the electron-phonon coupling.  Crucially, in the tight-binding and low-energy methods, the additional terms enable the recovery of the all-electron frequency-dependent vibrational responses, as quantified by the fulfilment of frequency-dependent vibrational sum rules relating them to the electromagnetic susceptibilities. Therefore, the correction changes qualitatively the tight-binding vibrational responses, which would otherwise have a zero sum rule. For instance, in metallic gapped graphene, the nuclear velocity contribution leads to an excellent agreement with first-principles calculations of the sum rule for the Born effective charges. Moreover, for high doping levels in this system, the tight-binding correction to the Born effective charges is quantitatively relevant, up to $\sim 50\%$. The changes in the Born effective charges are also affected by the electronic topology, as shown for the topologically non-trivial Haldane model. In general, the corrections to the vibrational responses are enhanced for a weak electron-phonon coupling.

Fundamentally, our equations do not simply provide a quantitative correction but they change the tight-binding vibrational responses from a qualitative viewpoint. In addition to the effects studied within this paper, the nuclear velocity corrections enable tight-binding models to capture phenomena such as vibrational circular dichroism. Furthermore,  we speculate that they can be relevant in the assessment of the properties related to chiral phonons \cite{Juraschek2025}, especially the phonon magnetic moment \cite{PhysRevLett.127.186403, PhysRevB.105.094305}.

\section*{Acknowledgements}
We acknowledge the MORE-TEM ERC-SYN project, Grant Agreement No. 951215. PF also acknowledges the funding from the project Ateneo 2025 by Sapienza - University of Rome (grant code: B83C25004300005). We acknowledge the ISCRA C projects (grant codes: HP10CRRY27 and HP10CMBQLG) by CINECA for the first-principles calculations. We thank Guglielmo Marchese and Stefano Paolo Villani for their support in the first-principles calculations. 
We thank Massimiliano Stengel, Raffaele Resta, Antimo Marrazzo and Giorgio Sangiovanni for useful discussions and suggestions.

\clearpage
\appendix
\begin{widetext}
\section{Details on Eq. \eqref{eq:HLCAORDOT} of the main text}\label{app:1}
Here, we derive in detail the expression for the $\mathcal{H}^{\mathrm{R,\dot{R}}}$ given in Eq. \eqref{eq:HLCAORDOT} of the main text. We remind that the temporal derivative of the velocity including atomic orbital, given in Eq. \ref{eq:derivative_vi_atomic_orbital} of the main text, is
\begin{equation}
\begin{aligned}
\frac{d\ket{\phi^{\dot{\mathbf{R}}_s,\mathbf{R}_{s}}_{si}}}{dt}=e^{i\alpha_s(\hat{\mathbf{r}})}\frac{i}{\hbar}\Bigg(m\ddot{\mathbf{R}}_s(t)\cdot(\hat{\mathbf{r}}-\mathbf{R}_s(t))-\dot{\mathbf{R}}_s(t)\cdot \hat{\mathbf{p}}-m|\dot{\mathbf{R}}_s(t)|^2\Bigg) 
\ket{\phi^{\mathbf{R}_{s}(t)}_{si}}.
\end{aligned} 
\end{equation}
The LCAO effective Hamiltonian with nuclear velocity dependent atomic orbitals (Eq. \eqref{eq:LCAO_Hamiltonian_R} of the main text) is
\begin{align}
&\mathcal{H}^{\mathrm{R,\dot{R}}}_{bj,b'i}=\mathcal{H}^{\mathrm{LCAO}}_{bj,b'i}
+\mathcal{M}_{bj,b'i}^{\rm R, \dot{R}}, 
\end{align}
\begin{align}
\mathcal{H}^{\mathrm{LCAO}}_{bj,b'i}=\braket{\phi^{\dot{\mathbf{R}}_b,\mathbf{R}_{b}}_{bj}|H^{\mathrm{AE}}|\phi^{\dot{\mathbf{R}}_{b'},\mathbf{R}_{b'}}_{b'i}}\\
\mathcal{M}_{bj,b'i}^{\rm R, \dot{R}}=\frac{i\hbar}{2} \Bigg(\frac{d\bra{\phi^{\dot{\mathbf{R}}_b,\mathbf{R}_{b}}_{bj}}}{dt}\ket{\phi^{\dot{\mathbf{R}}_{b'},\mathbf{R}_{b'}}_{b'i} }- \bra{\phi^{\dot{\mathbf{R}}_b,\mathbf{R}_{b}}_{bj}}\frac{d\ket{\phi^{\dot{\mathbf{R}}_{b'},\mathbf{R}_{b'}}_{b'i}}}{dt}\Bigg).
\end{align}
Expliciting the Hamiltonian in terms of the translated atomic orbitals $\{\ket{\phi^{\mathbf{R}_{s}(t)}_{si}}\}$, we have that, using the commutation relations between the position and momentum operators implying that $e^{-i\alpha_{s}(\hat{\mathbf{r}})}\hat{\mathbf{p}}e^{i\alpha_{s}(\hat{\mathbf{r}})}=\hat{\mathbf{p}}+m\dot{\mathbf{R}}_s(t)$,
\begin{align}
    \mathcal{H}^{\mathrm{R,\dot{R}}}_{bj,b'i}=           \bra{\phi^{\mathbf{R}_{b}(t)}_{bj}}\Bigg(e^{i(\alpha_{b'}(\hat{\mathbf{r}})-\alpha_{b}(\hat{\mathbf{r}}))}\left(\frac{(\hat{\mathbf{p}}+m\dot{\mathbf{R}}_{b'}(t))^2}{4m}+\frac{V(\hat{\mathbf{r}})}{2}\right)
   +\left(\frac{(\hat{\mathbf{p}}+m\dot{\mathbf{R}}_{b}(t))^2}{4m}+\frac{V(\hat{\mathbf{r}})}{2}\right)e^{i(\alpha_{b'}(\hat{\mathbf{r}})-\alpha_{b}(\hat{\mathbf{r}}))}\Bigg)\ket{\phi^{\mathbf{R}_{b'}(t)}_{b'i}},
\end{align}

\begin{align}
\mathcal{M}_{bj,b'i}^{\rm R, \dot{R}}=\frac{i\hbar}{2} \Bigg(\frac{d\bra{\phi^{\dot{\mathbf{R}}_b,\mathbf{R}_{b}}_{bj}}}{dt}\ket{\phi^{\dot{\mathbf{R}}_{b'},\mathbf{R}_{b'}}_{b'i} }- \bra{\phi^{\dot{\mathbf{R}}_b,\mathbf{R}_{b}}_{bj}}\frac{d\ket{\phi^{\dot{\mathbf{R}}_{b'},\mathbf{R}_{b'}}_{b'i}}}{dt}\Bigg)\\
=\frac{1}{2} \bra{\phi^{\mathbf{R}_{b}(t)}_{bj}}\left(m\ddot{\mathbf{R}}_{b}(t)\cdot(\hat{\mathbf{r}}-\mathbf{R}_{b}(t))-\dot{\mathbf{R}}_{b}(t) \cdot \hat{\mathbf{p}}-m|\dot{\mathbf{R}}_{b}(t)|^2\right)e^{i\alpha_{b'}(\hat{\mathbf{r}})-i\alpha_{b}(\hat{\mathbf{r}})}\ket{\phi^{\mathbf{R}_{b'}(t)}_{b'i} }\\
+\frac{1}{2} \bra{\phi^{\mathbf{R}_{b}(t)}_{bj}}e^{i\alpha_{b'}(\hat{\mathbf{r}})-i\alpha_{b}(\hat{\mathbf{r}})}\left(m\ddot{\mathbf{R}}_{b'}(t)\cdot(\hat{\mathbf{r}}-\mathbf{R}_{b'}(t))-\dot{\mathbf{R}}_{b'}(t)\cdot \hat{\mathbf{p}}-m|\dot{\mathbf{R}}_{b'}(t)|^2\right)\ket{\phi^{\mathbf{R}_{b'}(t)}_{b'i} }
\end{align}
Summing the two contributions, the terms linear in the nuclear velocity are cancelled, yielding Eq. \eqref{eq:HLCAORDOT} of the main text,
\begin{align}
 &\mathcal{H}^{\mathrm{R,\dot{R}}}_{bj,b'i}=           \bra{\phi^{\mathbf{R}_{b}(t)}_{bj}}\Bigg(e^{i(\alpha_{b'}(\hat{\mathbf{r}})-\alpha_{b}(\hat{\mathbf{r}}))}\left(\frac{\hat{\mathbf{p}}^2}{4m}+\frac{V(\hat{\mathbf{r}})}{2}\right)
   +\left(\frac{\hat{\mathbf{p}}^2}{4m}+\frac{V(\hat{\mathbf{r}})}{2}\right)e^{i(\alpha_{b'}(\hat{\mathbf{r}})-\alpha_{b}(\hat{\mathbf{r}}))}\Bigg)\ket{\phi^{\mathbf{R}_{b'}(t)}_{b'i}}\nonumber\\
  &+ \frac{1}{2}\bra{\phi^{\mathbf{R}_{b}(t)}_{bj}}\Bigg(m \ddot{\mathbf{R}}_{b'}(t)\cdot\left(\hat{\mathbf{r}}-\mathbf{R}_{b'}(t)\right)+m\ddot{\mathbf{R}}_{b}(t)\cdot\left(\hat{\mathbf{r}}-\mathbf{R}_{b}(t)\right)
-\frac{m}{2}\Big(|\dot{\mathbf{R}}_{b'}(t)|^2+|\dot{\mathbf{R}}_{b}(t)|^2\Big)\Bigg)e^{i(\alpha_{b'}(\hat{\mathbf{r}})-\alpha_{b}(\hat{\mathbf{r}}))}\ket{\phi^{\mathbf{R}_{b'}(t)}_{b'i}}.
\label{eq:HLCAORDOT_app}
\end{align}

\end{widetext}

\bibliography{biblio}

\end{document}